\documentclass[fleqn,usenatbib]{mnras}
\usepackage{newtxtext,newtxmath}
\usepackage[T1]{fontenc}
\DeclareRobustCommand{\VAN}[3]{#2}
\let\VANthebibliography\thebibliography
\def\thebibliography{\DeclareRobustCommand{\VAN}[3]{##3}\VANthebibliography}

\usepackage{graphicx}	
\usepackage{amsmath}	
\usepackage{gensymb}
\usepackage{booktabs}
\usepackage{subcaption}
\usepackage{ragged2e}
\usepackage{float}
\usepackage{lscape}
\usepackage{xcolor}
\usepackage{titlesec}
\usepackage{adjustbox}
\usepackage{placeins}
\usepackage{wrapfig}
\titleformat{\subsubsection}{\normalfont\small\bfseries\upshape}{\thesubsubsection}{1em}{}
\titlespacing{\section}{0pt}{24pt}{-\parskip}
\titlespacing{\subsection}{0pt}{10pt}{-\parskip}
\titlespacing{\subsubsection}{0pt}{10pt}{-\parskip}
\newcommand{\dashedrule}{\leaders\hbox to 5pt{\hss.\hss}\hfill}

\title[Weak secondary cyclotron line in Cen X-3]{Weak secondary cyclotron line in eclipsing High Mass X-ray Binary Cen X-3}

\author[Dangal,P. et al.]{
Pravat Dangal,$^{1,2}$\thanks{E-mail: astro.dangal@gmail.com}
Ranjeev Misra,$^{3}$
Nand Kumar Chakradhari,$^{1,4}$
and Yashpal Bhulla$^{5}$
\\
$^{1}$School of Studies in Physics and Astrophysics, Pt. Ravishankar Shukla University, Raipur 492010, India\\
$^{2}$St. Joseph's College, Darjeeling 734104, India\\
$^{3}$Inter-University Centre for Astronomy and Astrophysics, Ganeshkhind, Pune 411007, India\\
$^{4}$Centre for Mega Projects in Multiwavelength Astronomy, Pt. Ravishankar Shukla University, Raipur 492010, India\\
$^{5}$Pacific Academy of Higher Education and Research University, Udaipur 313003, India
}

\date{Accepted XXX. Received YYY; in original form ZZZ}

\pubyear{2023}

\begin{document}
\label{firstpage}
\pagerange{\pageref{firstpage}--\pageref{lastpage}}
\maketitle

\begin{abstract}
We report the time resolved spectroscopy result from two observations of Cen X-3, over one binary orbit with ASTROSAT and two binary orbit with NuSTAR. NuSTAR covered two intensity states where the light curve showed transition in count rate from first to second binary orbit by a factor of $\sim$ 3. A phenomenological model comprising of partially absorbed powerlaw with smoothed high energy cutoff, cyclotron absorption $\sim$ 24 keV and 6.4 keV iron emission gave good fit for ASTROSAT observation. NuSTAR spectra required two additional emission components, a broad one $\sim$ 5.7 keV and a narrow one $\sim$ 6.9 keV. A weak secondary absorption feature at $\sim$ 11.6 keV and $\sim$ 14.5 keV was seen in the residuals of the spectral fit for ASTROSAT and NuSTAR data respectively. The secondary absorption energy showed no correlation with the cutoff energy. Its strength varied within 0.1 to 0.6 keV with its width $\sim$ 1.6 keV. Its energy and optical depth showed linear positive correlation with the fundamental cyclotron line energy and depth respectively. The cyclotron line energy showed anti-correlation to flux described by a powerlaw with negative index and the secondary absorption also showed similar trend to flux. Depth of secondary absorption was $\sim$ 45 \% and centroid energy was $\sim$ 54 \% of fundamental. Depth and energy ratio of secondary to fundamental lied within 2$\sigma$ deviation from 0.5. We suggest this secondary absorption to be a redshifted dipolar cyclotron resonance feature exhibiting sub-harmonic behaviour.

\end{abstract}



\section{Introduction}

Among the most luminous objects in the X-ray regime are the accretion-powered X-ray binaries (XRBs), which came to discovery in the 1970s \citep{giac71,tana72}, Centaurus X-3 (Cen X-3 hereafter) being first of its kind, by UHURU. The host star accretes matter from the companion star in two basic modes: disk accretion via Roche Lobe overflow and wind accretion. In accretion-powered Pulsars, the matter falls onto the neutron star's surface, forming an accretion column \citep{beck05}. High-Mass X-ray Binaries (HMXBs) have young optical companion stars of spectral type B or O, typically with mass M $\ge$ 10 M$_{\odot}$ and high magnetic fields B $\sim$ 10$^{12}$ G. Low-Mass X-ray Binaries (LMXBs) have older optical companions with mass M $\le$1 M$_{\odot}$ and lower magnetic fields B $\sim$ 10$^{9}$ to 10$^{10}$ G.

Cen X-3 is a widely studied HMXB, which hosts a neutron star and a young and massive O6-8 type supergiant star (V779 CEN or Krzeminski's star) as the optical component \citep{krzemenski1974}. The mass of the optical star is estimated to be M$_{c}$  $\sim$ 20.5 M$_{\odot}$ and radius to be R$_{c}$  $\sim$ 12 R$_{\odot}$ \citep{Hutchings1979}. Distance to the binary system was estimated earlier to be $\sim$ 8 kpc \citep{krzemenski1974} and later to be 5.7  $\pm$  1.5 kpc \citep{thom09} and recent to be 7.2 kpc \citep{vall22}. This eclipsing HMXB is known to pulsate at $\sim$ 4.8 s, and has an orbital period $\sim$ 2.1 days \citep{schreier1972evidence}. Cen X-3 is one of the brightest sources in X-rays with a luminosity of $\sim$ 5 $\times$ 10$^{37}$ erg\,s$^{-1}$ \citep{suchy2008}. It is a disk-fed system by Roche Lobe overflow. A QPO of 35  $\pm$  2 mHz was measured by \citet{take91}, $\sim$ 40 mHz by \citet{raichur2008} and recently $\sim$ 26 mHz in the low state by \citet{bachhar2022}, thereby supporting the evidence of accretion disk. The compact neutron star mass M$_{n}$ is estimated to be 1.2  $\pm$  0.21 M$_{\odot}$ and has an inclination of 70$\degree$ with the plane of the binary orbit  \citep{asht99}.

Initial estimation of neutron star magnetic field $\sim$  2.6 $\times$ 10$^{12}$ G using quasi-periodic oscillation analysis at frequency $\sim$ 40 mHz was reported with Broad Band X-Ray Telescope (BBXRT) observation of 1990, December 9 \citep{audley98}. The same work showed evidence of cyclotron resonance scattering feature (CRSF hereafter) in Cen X-3 with energy ($E_{cyc}$ hereafter) at 25.1 $\pm$ 0.3 keV with RXTE observation of 1996, March 1. Using 45.36 ks BeppoSAX observation of 1997, February 27 and 28, \citet{santa98} reported the X-ray spectrum to be described by an absorbed power law model with a high energy cutoff $\sim$ 14 keV folded around 8 keV along with a soft excess $\sim$ 0.1 keV, one iron emission line $\sim$ 6.7 keV and a CRSF with energy $\sim$ 28 keV at high significance. Lines at  6.4, 6.7 and 6.9 keV were reported earlier from ASCA observation of 1995, February 1 in the eclipse phase of the source by \citet{ebis96} and later with Suzaku observation of 2008, December 8  by \citet{Naik}. The 6.4 keV line is known to be associated with the fluorescence from the cold matter at close proximity to the neutron star \citep{naga89}, and 6.7 and 6.9 keV lines originating from accretion disk corona \citep{killman1989}, which is highly ionized. 

Ever since the detection of CRSF in Cen X-3 by BeppoSAX, this feature has been detected by later missions and its $E_{cyc}$ is widely studied for pulse phase dependency and correlations with other spectral parameters. In BeppoSAX observation, pulse phase resolved analysis showed asymmetric variation of $E_{cyc}$ decreasing from ascent to descent in the pulse profile \citep{santa98,burderi2000}. 

The X-ray spectrum of an XRB is typically described by a high energy cutoff powerlaw, with cutoff energy $E_{cut}$ $\sim$ kT$_e$, where T$_e$ is the electron temperature. This is an important parameter that provides temperature of the surrounding plasma. Electrons gain energy by Compton down-scattering photons with energy above kT$_e$ thereby leading high-energy photons to lose energy to the electrons. This result in a cutoff in the X-ray spectrum at higher energies, as fewer photons are able to escape with high energies due to repeated scattering events. Correlation of $E_{cyc}$ with $E_{cut}$ facilitates study of cyclotron line dependence upon temperature of the surrounding plasma. In an attempt to study correlation of fundamental $E_{cyc}$ with $E_{cut}$, \cite{staubert2019} sampled different sources reported to have CRSF (Cen X-3 being one) and found a positive correlation where cutoff energy was roughly proportional to square root of $E_{cyc}$. However, correlation study of $E_{cyc}$ and $E_{cut}$ has not been done individually for Cen X-3. 

Correlation of $E_{cyc}$ with luminosity for many X-ray binary systems has been studied previously. Some sources showed positive correlation, some showed negative correlation and for some sources correlation with luminosity has not been observed \citep[][Table 7]{staubert2019}. Cen X-3 has been studied for its $E_{cyc}$ dependency to luminosity, sampling separate observations with short exposures and no correlation has been seen yet. Most recent study of $E_{cyc}$ with luminosity for multiple states of Cen X-3 with previous four ASTROSAT observations has been done by \cite{bachhar2022} and they found no dependence of $E_{cyc}$ to luminosity. 

Different values of $E_{cyc}$ have been reported for different periods of observation. Measurements of $E_{cyc}$ with previous missions are enlisted by \citet[][Table 4]{tomar2021}, that shows variations in line energy between 27.9  $\pm$  0.5 keV and $30.7^{+0.5}_{-0.4}$ keV. They have reported new measurements of line energy at $29.22^{+0.28}_{-0.27}$ keV with NuSTAR observation of 2015, November 30 and at $30.29^{+0.68}_{-0.61}$ keV with Suzaku observation of 2008, December 8. Two Insight-HXMT observations, of 25 ks and 6.7 ks exposure, have reported $E_{cyc}$ at $27.39^{+2.67}_{-0.3}$ and $26.29^{+0.78}_{-0.83}$ keV respectively along with discovery of a CRSF harmonic at $\sim$ 47 keV \citep{wang2023}. Also, pulse phase resolved analysis has revealed its $E_{cyc}$ to fluctuate by $\sim$ 30\% with mean at 28 keV \citep{staubert2019}. We can infer from above that cyclotron absorption feature in Cen X-3 shows variability in its centroid energy within a single pulse as well as at different intensity states where many pulses are averaged. A time resolved analysis would be crucial in studying variability of CRSF at different orbital phase. Pulse-phase resolved analysis at different orbital-phases would further provide opportunity to investigate the dependence of $E_{cyc}$ with pulse at various flux levels.

In works of \cite{burderi2000}, with BeppoSAX, a faint absorption near cutoff energy was seen in the spectral fit residuals and also in the previous observations from ASCA $\sim$ 10 keV \citep{naga89} and with GINGA. \citet{burderi2000} interpreted this dip as feature of the \texttt{highecut} model, since the discontinuity at cutoff energy may develop an artifact cusp, apparently showing absorption like feature. They modified the \texttt{highecut} model with a smoothing width at cutoff energy. Later, \cite{Naik} conducted a time resolved spectroscopy of Cen X-3 with Suzaku observation of 2008, which covered a full orbital period (2.1 days). This faint absorption was not seen in their spectral fit and the authors restricted their analysis to study the iron K$\alpha$ emission lines properties over orbital phase. CRSF and its correlations with other spectral parameters was not explored. 

In this study, we have carried out a detailed time-resolved spectral analysis of the source averaged for pulse-phase with SXT and LAXPC onboard ASTROSAT with observation exposure of 72.953 ks and with longest NuSTAR observation with exposure of $\sim$ 189 ks in two different intensity states in two binary orbits covered continuously in single observation. Both the observation's analysis has been done for the first time. We have laid emphasis on CRSF absorption of the source at different flux levels.

In section ~\ref{sec:obs}, details of observation and data reduction has been laid down. In Section ~\ref{sec:results} analysis and results is provided. In section ~\ref{sec:discuss} and section ~\ref{sec:conclusion} we have discussions and conclusion.  

\begin{figure*}
    \resizebox{17cm}{6cm}{\includegraphics[width=\textwidth]{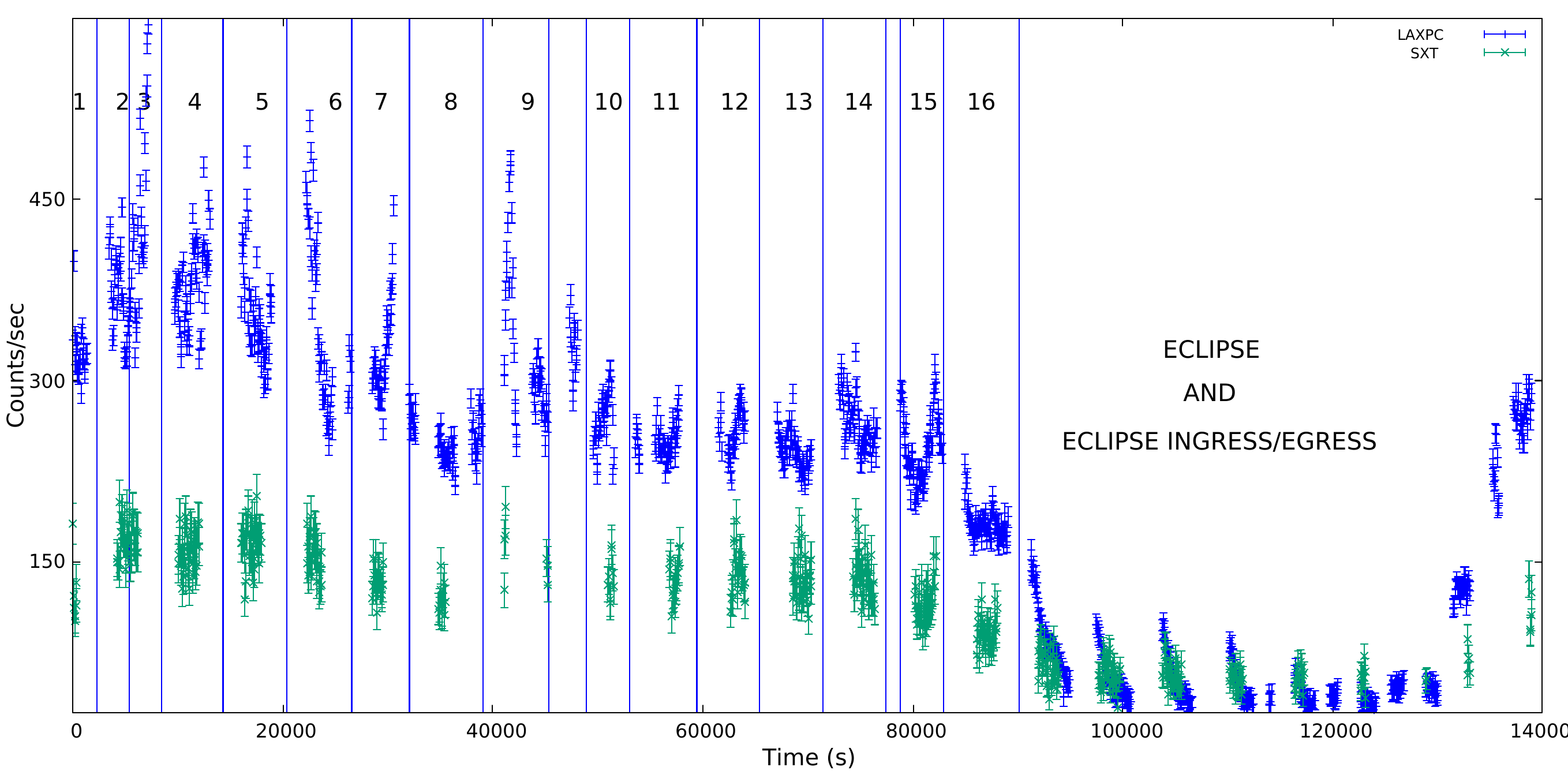}}
    \caption{Segmentation of light curve from LAXPC and SXT for Cen X-3. Segments 1 to 9 are in the high-intensity state with high variability, segment 8 has a dip and segment 9 has a bump, segments 10 to 16 have almost uniform count rates. Segments beyond 16 are eclipse ingress, eclipse and eclipse egress. Synchronous data from both detectors have been extracted from each segment for joint spectral fitting. Segments where SXT data could not be enclosed with LAXPC has been excluded for analysis. Note: The low SXT count rate has been scaled up in the figure by a factor for joint representation with a higher LAXPC count rate.}
    \label{fig:astro_seg}
\end{figure*}
\section{Observations and Data Reduction}
\label{sec:obs}
\vspace{5pt}
As India's first multi-wavelength astronomy satellite, ASTROSAT \citep{kpsingh2014} was launched into a low orbit of 650 km on 2015, September 28. This mission carries four active payloads individually sensitive in broad energy band ranging from optical to hard X-rays. Ultra Violet Imaging Telescope (UVIT) sensitive to optical/UV regime, Soft X-ray Telescope (SXT) for soft X-rays from 0.3 to 8 keV, Large Area X-ray Proportional Counter (LAXPC) in hard X-rays from 3 to 80 keV and Cadmium Zinc Telluride Imager (CZTI) in hard X-rays from 10 to 100 keV. 

SXT is an imaging telescope that has charge coupled device (CCD). Its effective area is $\sim$ 90 cm$^2$ at 1.5 keV with instrument PSF of $\sim$ 3 arcmin. It has time resolution of 0.278 s with 150 $\times$ 150 centred pixel frame \citep{singh2017soft}. LAXPC is a non imaging instrument that has three identical co-aligned units LAXPC10, LAXPC20 and LAXPC30 which has by far the largest effective area of $\sim$ 6000 cm$^2$ within 5 to 20 keV energy range and have a time resolution of 10 $\micro s$ which makes it ideal for timing and spectral analysis \citep{yadav2016}.

Launched in 2012, June 13, NuSTAR \citep{harrison2013}, is an imaging telescope, effective in the energy range 3.0-79.0 keV. It consists of two co-aligned units that focus X-ray photons
on to their respective Focal Plane Modules (FPMA and FPMB). Each module consists of Cadmium Zinc Telluride pixel detectors. These modules individually provides energy resolution of 0.4 keV at 10 keV and 0.9 keV at 60 keV (FWHM) making them ideal for spectral analysis. Each unit has 32 $\times$ 32, 0.6-mm pixels providing a 12 arcmin field of view with an angular resolution of 18 arcsec (FWHM).
\subsection{Observation}
\vspace{5pt}
Details of Cen X-3 observations from ASTROSAT and NuSTAR are provided in Table ~\ref{tab:observations}.
\subsubsection{ASTROSAT Observation}
We have used Obs ID A04\_216T01\_9000001968 from 2018, March 8 to 10 having simultaneous observations from LAXPC and SXT consisting of twenty mission orbits covering one complete binary orbit with an exposure of 72.953 ks. This observation has high count rate states, intermediate states, dips, eclipse ingress, eclipse and eclipse egress.

\subsubsection{NuSTAR Observation}
\vspace{5pt}
We have used Obs ID 30701019002 from 2022, January 12 to 16 which is the most latest and the longest ever observation of the source with an exposure of $\sim$ 189 ks covering almost two binary orbit of the source in two different intensity states.

Both ASTROSAT and NuSTAR observations are in low (hard) state in comparison to previous observations. Average count rates of detectors are provided in Table ~\ref{tab:observations}.\\

\begin{table}
\small
\setlength{\tabcolsep}{2.5pt}
    \centering
    \begin{tabular}{@{}lcccc@{}}
        \toprule
        \textbf{Mission}&\textbf{Obs ID} & \textbf{Start date} & \textbf{Exposure} & \textbf{MJD} \\
        \midrule
        &&&&\\
        ASTROSAT & A04\_216T01\_9000001968 & 2018-03-08 & 72.953 ks & 58185.65 \\
        &&&&\\
        NuSTAR & 30701019002 & 2022-01-12 & 189 ks & 59591.12 \\
        &&&&\\
        \midrule
        \textbf{Payloads}&\multicolumn{4}{c}{\textbf{Average count rate (cts/sec)}}\\
        \midrule
        LAXPC & \multicolumn{4}{c}{597.6}\\
        SXT & \multicolumn{4}{c}{1.13}\\
        FPMA & \multicolumn{4}{c}{70.5}\\
        FPMB & \multicolumn{4}{c}{69.6}\\
        \bottomrule
    \end{tabular}
    \caption{Observation log of ASTROSAT and NuSTAR data}
    \label{tab:observations}
\end{table}

\subsection{Data Reduction}
\vspace{5pt}
Details of data reduction tools and techniques are provided below for detectors of both the missions.
\subsubsection{ASTROSAT}
\vspace{5pt}
We have used the standard data reduction tools (Format A)\footnote{\url{http://astrosat-ssc.iucaa.in/laxpcData}} for extracting LAXPC products. The Level1 data from ASTROSAT archive was used to extract light curve and spectrum files. The good time interval (hereafter GTI) was created with \texttt{laxpc\_make\_stdgti} tool that carefully excludes the Earth occultation periods, South Atlantic Anomaly (SAA) passage, etc. Respective eventfile and filterfiles were created with $\texttt{laxpc\_make\_filelist}$. Level2 event fits file for all the orbits merged together, was created using $\texttt{laxpc\_make\_event}$. Finally we used $\texttt{laxpc\_make\_lightcurve}$, $\texttt{laxpc\_make\_spectra}$, $\texttt{laxpc\_make\_backspectra}$, $\texttt{laxpc\_make\_backlightcurve}$ to extract light curve, spectrum, background spectrum and background light curve respectively. We extracted light curve from LAXPC with binsize of 64 sec for full energy range 3 to 80 keV for primary analysis. Since LAXPC30 suffered gas leakage since 2017 which resulted in unreal gain, so we excluded data from LAXPC30. After analysing data from LAXPC10 we found that this unit too operated in high gain mode, so we implemented only LAXPC20 for our analysis. 

We have used standard SXT Level2 data from archive. We merged the cleaned event files from all orbits using \texttt{SXTPYJULIAMERGER}\footnote{\url{https://github.com/gulabd/SXTMerger.jl}} to create merged clean event fits file. We implemented \texttt{SAOImageDS9} to select the source and background regions. Finally this merged cleaned event file was taken as input in Heasoft's \texttt{xselect} package to extract SXT products. Since no significant pile up was found so we selected a circular source region of radius 17 arcmin centred at source's maximum intensity co-ordinates. We set binsize to minimum for SXT at 2.3775 s. Background and response for spectra has been used as provided by the SXT team. The ARF for present observation spectra was generated using $\texttt{sxt\_ARFModule\_v02}$ which took the recent versions of ARF provided by team as input. This tool scales the ARF as per the observation specific parameters like the area of the source extraction region and the distance of source centroid from the optical axis thereby providing correction for vignetting and writing a new ARF file.

For SXT data, we applied gain shift correction with \texttt{gain fit}. Slope was fixed at 1 allowing offset to vary initially. Then both the gain parameters were frozen for rest of the analysis. A systematic of 2 percent is applied for joint fitting of SXT and LAXPC data suggested by the instruments team. Gain shift correction and systematics are applied in the spectral fitting for all the segments.

\subsubsection{NuSTAR}
\vspace{5pt}
NuSTARDAS v2.1.1 integrated with the HEASOFT version 6.29 and CALDB version 20230124 has been used for the data reduction. We have used the cleaned event files for extracting the products using \texttt{nuproducts}. We have selected a circular region of 180 arcsec for source and same radius for background far away from source. Initially products were extracted for full exposure with both the focal planes. Light curve revealed high degree of variability in count rates with two intensity states in each of the orbital phase. Products were extracted using \texttt{nuproducts} using suitable GTI for each orbital phase segments for both detectors. NuSTAR's \texttt{nuproducts} take cares of respective response, ancillary response, background files in the extraction process. 

HEASOFT tool \texttt{lcmath} was used to create a background corrected light curve for all the detectors in both the missions. For both missions and each detectors onboard, time segments exposure was selected as a result of initial routine spectral inspections of parameter over series of initial small exposures, considering orbital phases with high counts, low counts, dips and bumps.

\section{Analysis and Results}
\label{sec:results}
\vspace{5pt}
High variability in count rate over the binary orbit of Cen X-3 led us to adopt time-resolved analysis. To achieve this, we have created GTI that divide the light curve into various segments which contains simultaneous data from LAXPC and SXT for ASTROSAT and subsequently from FPMA and FPMB for NuSTAR, making spectra suitable for joint fitting for each mission. We have made use of \texttt{XSPEC version:12.12.1c} for spectral analysis.\\

\subsection{ASTROSAT}
\vspace{5pt}
Segmentation of light curve from LAXPC and SXT for Cen X-3 is shown in Figure ~\ref{fig:astro_seg}. We have adopted energy range from 1 to 8 keV for SXT and 3 to 30 keV for LAXPC20. We have excluded eclipse ingress, eclipse and eclipse egress for our investigation. We initially implemented a phenomenological model to the spectra for the segments (Figure~\ref{fig:astro_seg}) that consists of a typical powerlaw with high energy cut-off $\sim$ 14 keV and folding energy $\sim$ 8 keV (\texttt{highecut}) with one gaussian component to account for line emission $\sim$ 6.4 keV, one gaussian absorption component (\texttt{gabs}) for cyclotron resonance feature and interstellar absorption. This gave reduced chi-squared ($\chi^{2}_{red}$) mostly above 2. We introduced a partially covering component in addition to the existing components to take into account any additional hydrogen column density close to the system. This provided a fit with $\chi^{2}_{red}$ $< 2$. We then extended this model to all segments, which reproduced acceptable fits with 1< $\chi^{2}_{red}$ < 1.5. Few parameters showed negligible variation over orbital phase. In order to investigate the influence of model parameters to the fit we implemented a technique for constraining parameters with error estimation of all parameters that was kept free initially. We fit the values within error bars with a constant function to probe the possibility to fix the respective parameter throughout the analysis. Results suggested Hydrogen column density for interstellar absorption $ nH1 \sim 0.625 \times 10^{22}$ units and line emission energy $\sim 6.43$ keV to be constant for all segments. We then fixed these parameters for further analysis.
\begin{figure}
\centering
   \resizebox{7cm}{5cm}{\includegraphics[width=\linewidth]{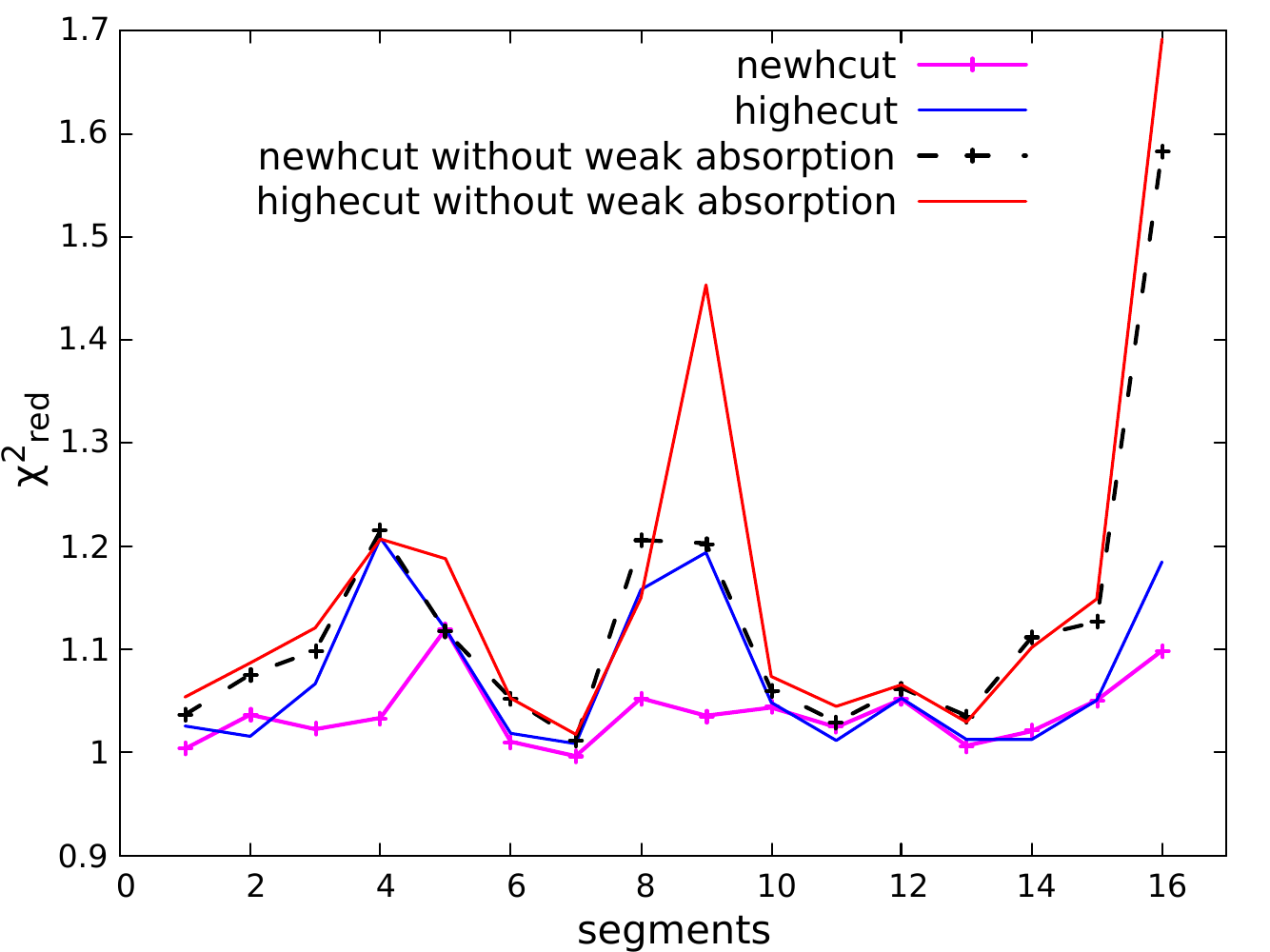}}
\caption{\small Comparison of $\chi^{2}_{red}$ using \texttt{newhcut} and \texttt{highecut} with and without additional absorption component in the continuum model.}
   \label{fig:astrochi}
\end{figure}

A faint absorption like feature around 10 keV to 12 keV could be sensed in the residuals. This feature was also seen in the works of \citet{burderi2000} with BeppoSAX data from 1997 February 27 19:45 to February 28 11:00 (UT), where it was interpreted as a consequence of \texttt{highecut} component in powerlaw that has discontinuous derivative at cutoff energy that results in a cusp thereby showing absorption like feature. They implemented a high energy cutoff powerlaw smoothed at the cutoff energy (now named as \texttt{newhcut}) to neutralise this feature. Best fit smoothing width was  3.1  $\pm$  0.3 keV. To investigate this, we adopted \texttt{newhcut} to ASTROSAT data with smoothing width fixed at 5 keV as folding energy ($E_{fold}$) was close to 8 keV. The absorption feature was still found to be present in almost all the segments. We then adopted an additional absorption component (\texttt{gabs}). This provided a very good fit to all segments $\chi^{2}_{red}$ $\sim$ 1. The comparison of the $\chi^{2}_{red}$ with additional absorption component and without absorption using \texttt{highecut} and \texttt{newhcut} is shown in Figure ~\ref{fig:astrochi}. It can be seen from the plot that the introduction of a weak absorption component in the model produced comparatively better fit for nearly all segments.
\begin{figure}
    \centering
    \begin{subfigure}[b]{0.43\textwidth}
         \centering
         \includegraphics[width=\textwidth]{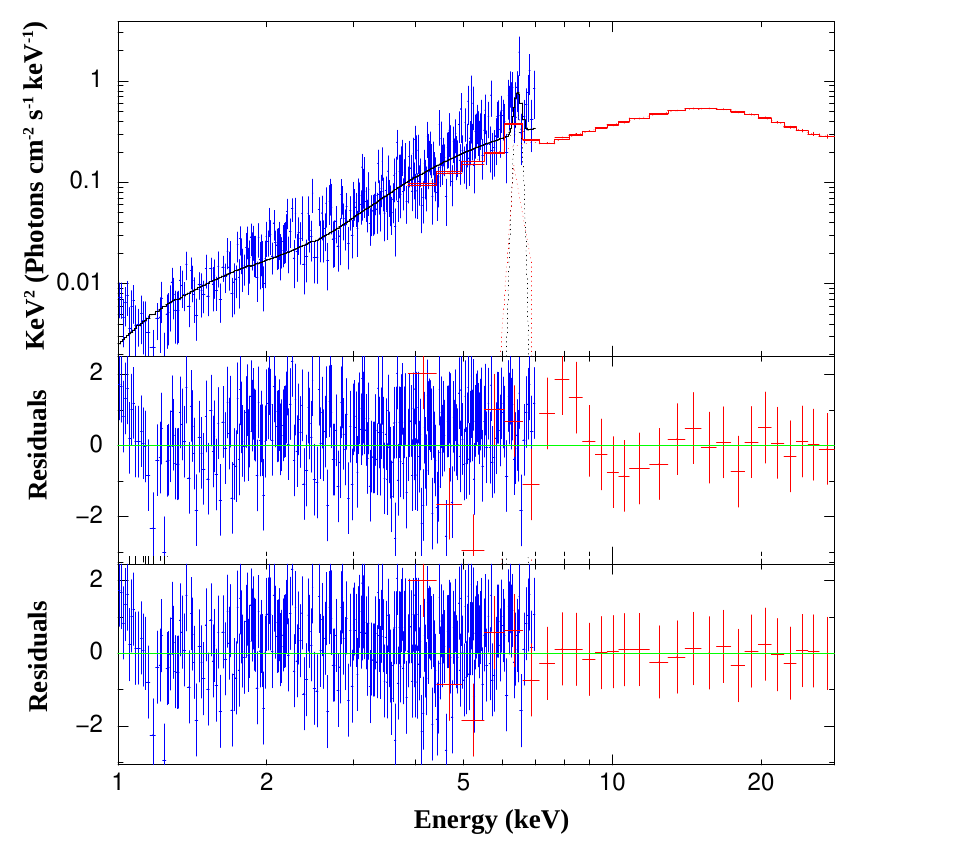}
         \caption{}
    \end{subfigure}
    \begin{subfigure}[b]{0.43\textwidth}
         \centering
         \includegraphics[width=\textwidth]{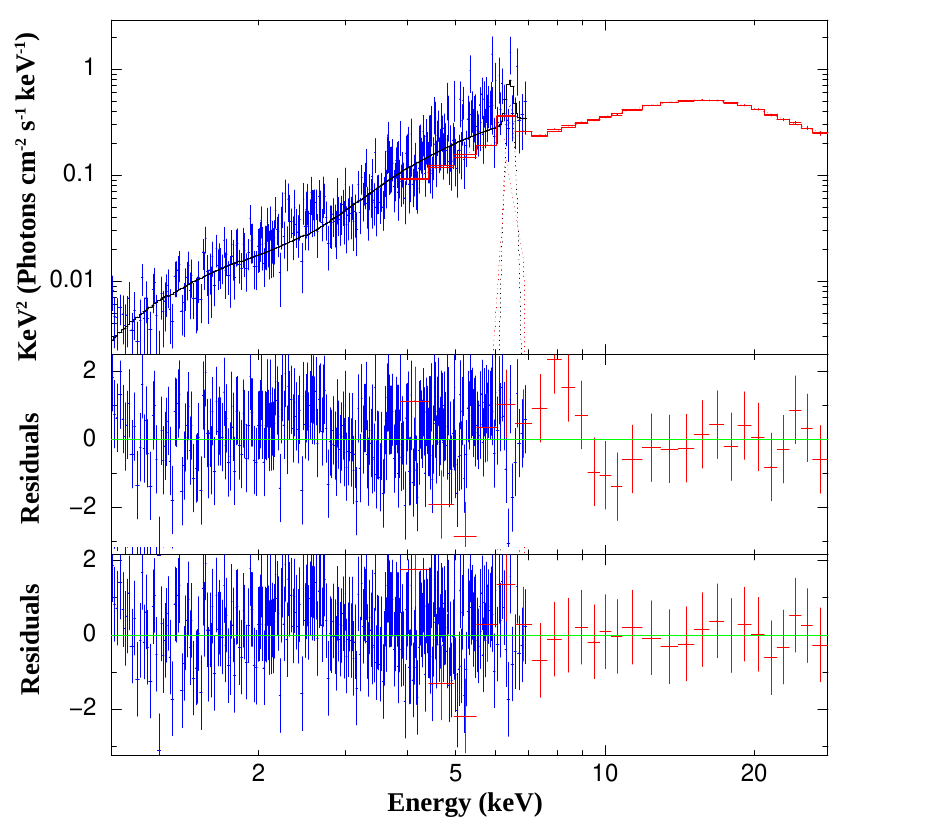}
         \caption{}
    \end{subfigure}
    \caption{\textbf{(a)}: Best fit spectra from joint fit of LAXPC and SXT observation onboard ASTROSAT for segment 4. \textbf{Upper panel}: Best fit unfolded spectra. \textbf{Middle panel}: Best fit residuals ((data-model)/error) for phenomenological model without weak absorption component. \textbf{Bottom panel}: Best fit residuals for phenomenological model with weak absorption component. \textbf{(b)}: Similar as (a) for segment 13. Segments 4 and 13 are from Figure ~\ref{fig:astro_seg}.}
    \label{fig:astrospec}
\end{figure}
\begin{figure}
    \centering
    \begin{subfigure}[b]{0.33\textwidth}
        \centering
            \includegraphics[trim=0.0cm 1.4cm 0.0cm 0.0cm, clip, width=\textwidth]{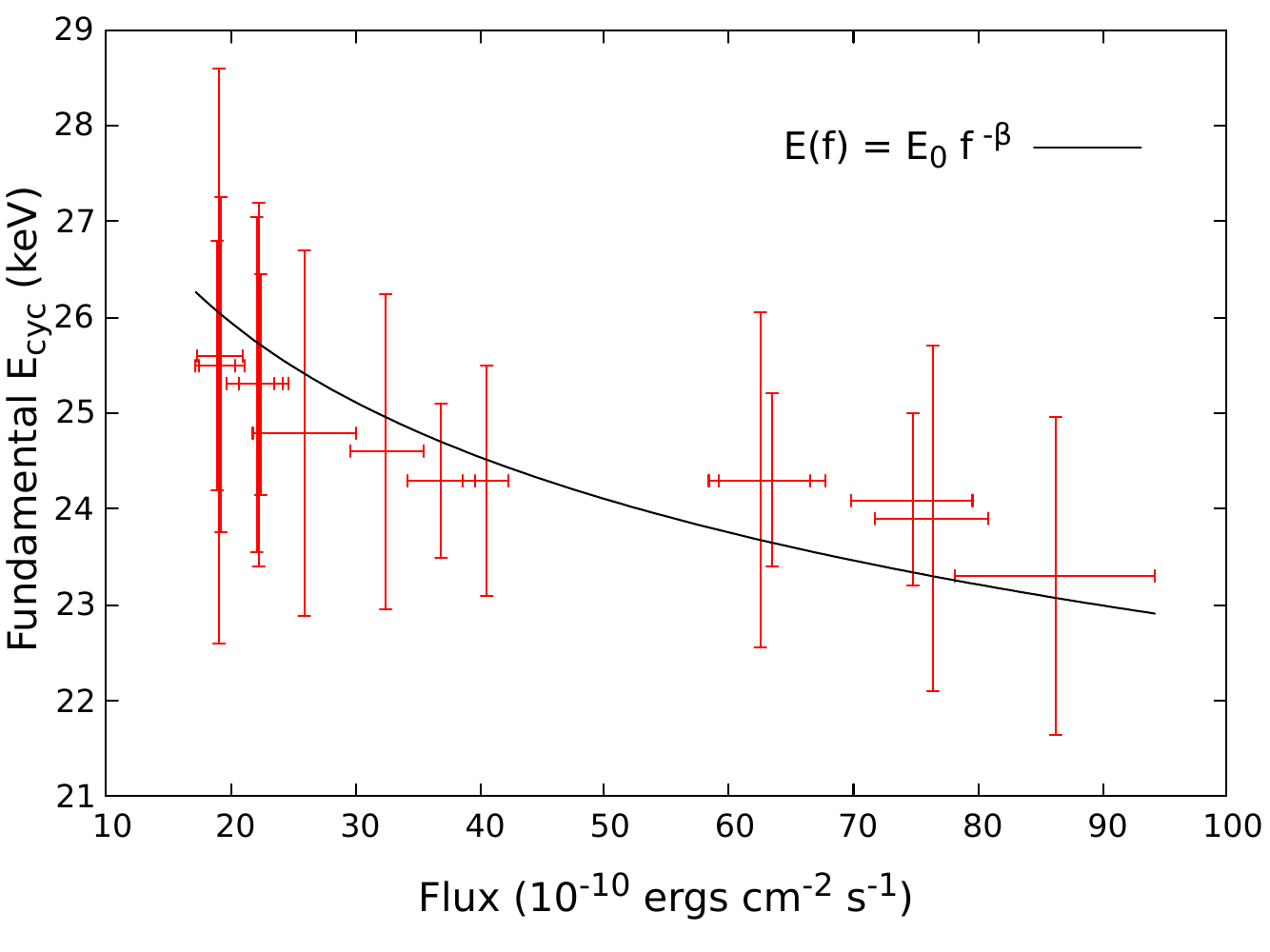} 
    \end{subfigure}
    \begin{subfigure}[b]{0.33\textwidth}
        \centering
        \includegraphics[trim=0.0cm 0.0cm 0.0cm 0.0cm, clip, width=\textwidth]{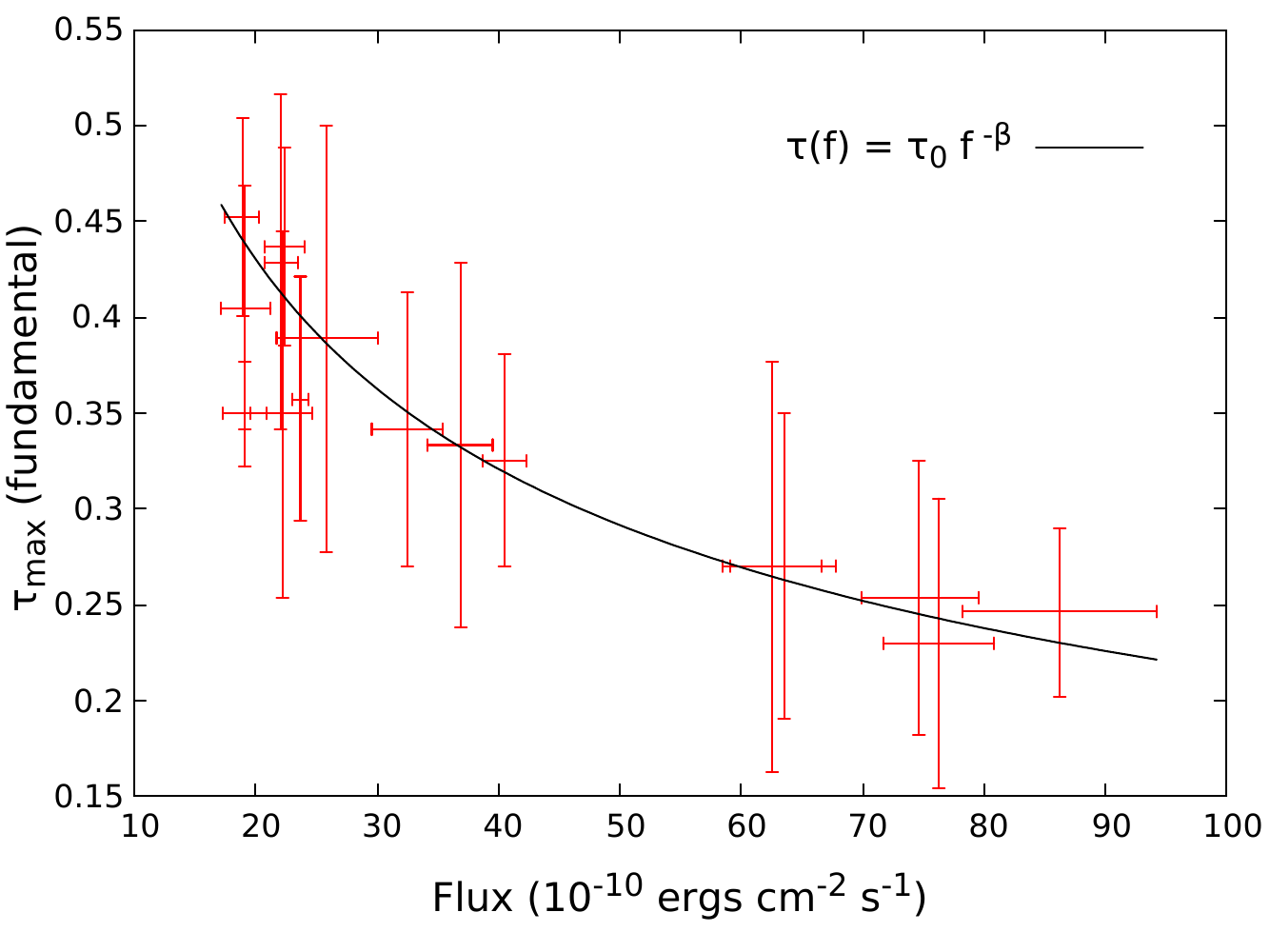}
    \end{subfigure}
    \caption{For ASTROSAT data, \textbf{Top}: Fundamental absorption energy vs 1 to 30 keV flux showing powerlaw anti-correlation trend with index $\beta$ = 0.08 $\pm$ 0.01 and normalisation E$_0$ = 33.1 $\pm$ 0.2 . \textbf{Bottom}: Powerlaw anti-correlation of index $\beta$ = 0.4 $\pm$ 0.01 and normalisation $\tau_0$ = 1.5 $\pm$ 0.04 between fundamental CRSF optical depth and flux.}
    \label{fig:astrofund}
 \end{figure}
 \begin{figure}
    \centering
    \begin{subfigure}[b]{0.35\textwidth}
        \centering
            \includegraphics[trim=0.0cm 1.4cm 0.0cm 0.0cm, clip, width=\textwidth]{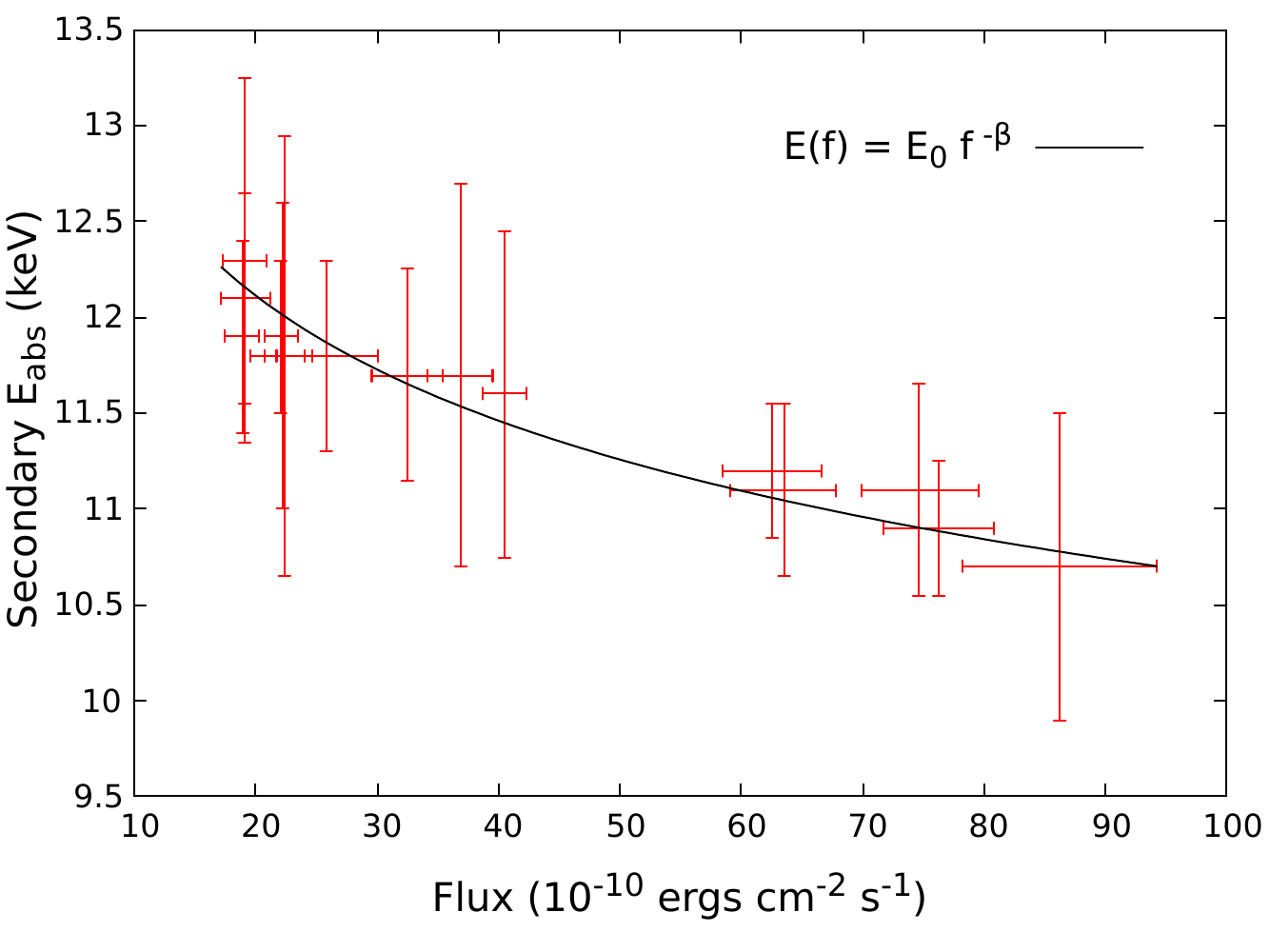} 
    \end{subfigure}
    \begin{subfigure}[b]{0.35\textwidth}
        \centering
        \includegraphics[trim=0.0cm 0.0cm 0.0cm 0.0cm, clip, width=\textwidth]{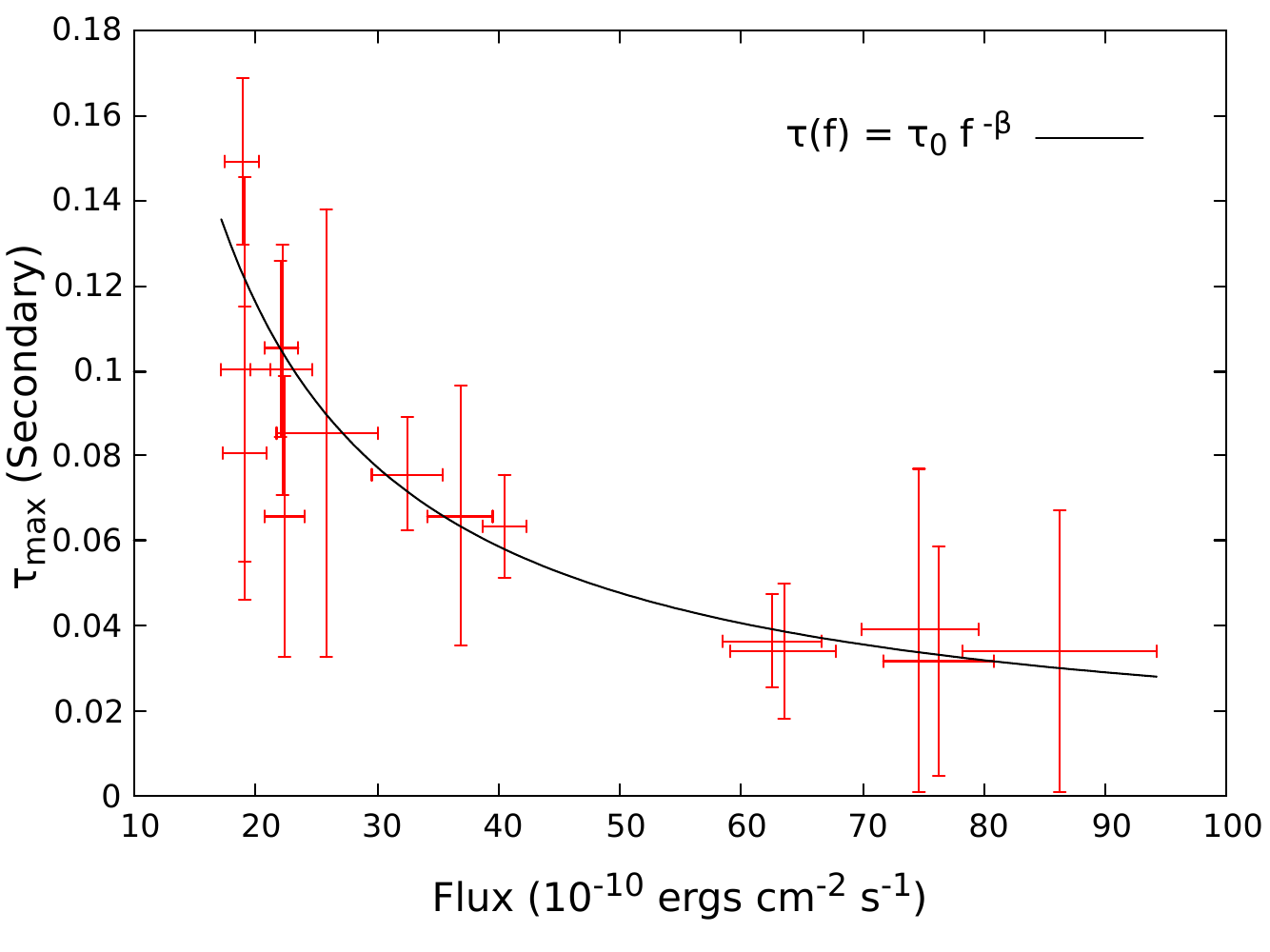}
    \end{subfigure}
    \caption{For ASTROSAT data, \textbf{Top}: Secondary absorption energy (E$_{\text{abs}}$) vs 1 to 30 keV flux showing powerlaw anti-correlation trend with index $\beta$ = 0.08 $\pm$ 0.01 and normalisation of E$_0$ = 15.4 $\pm$ 0.07. \textbf{Bottom}: Powerlaw anti-correlation of index $\beta$ = 1.1 $\pm$ 0.2 and normalisation $\tau_0$ = 2.9 $\pm$ 0.3 between secondary absorption optical depth and flux.}
    \label{fig:astroweak}
\end{figure}
\begin{figure}
    \centering
    \begin{subfigure}[b]{0.35\textwidth}
        \centering
            \includegraphics[trim=0.0cm 0.0cm 0.0cm 0.0cm, clip, width=\textwidth]{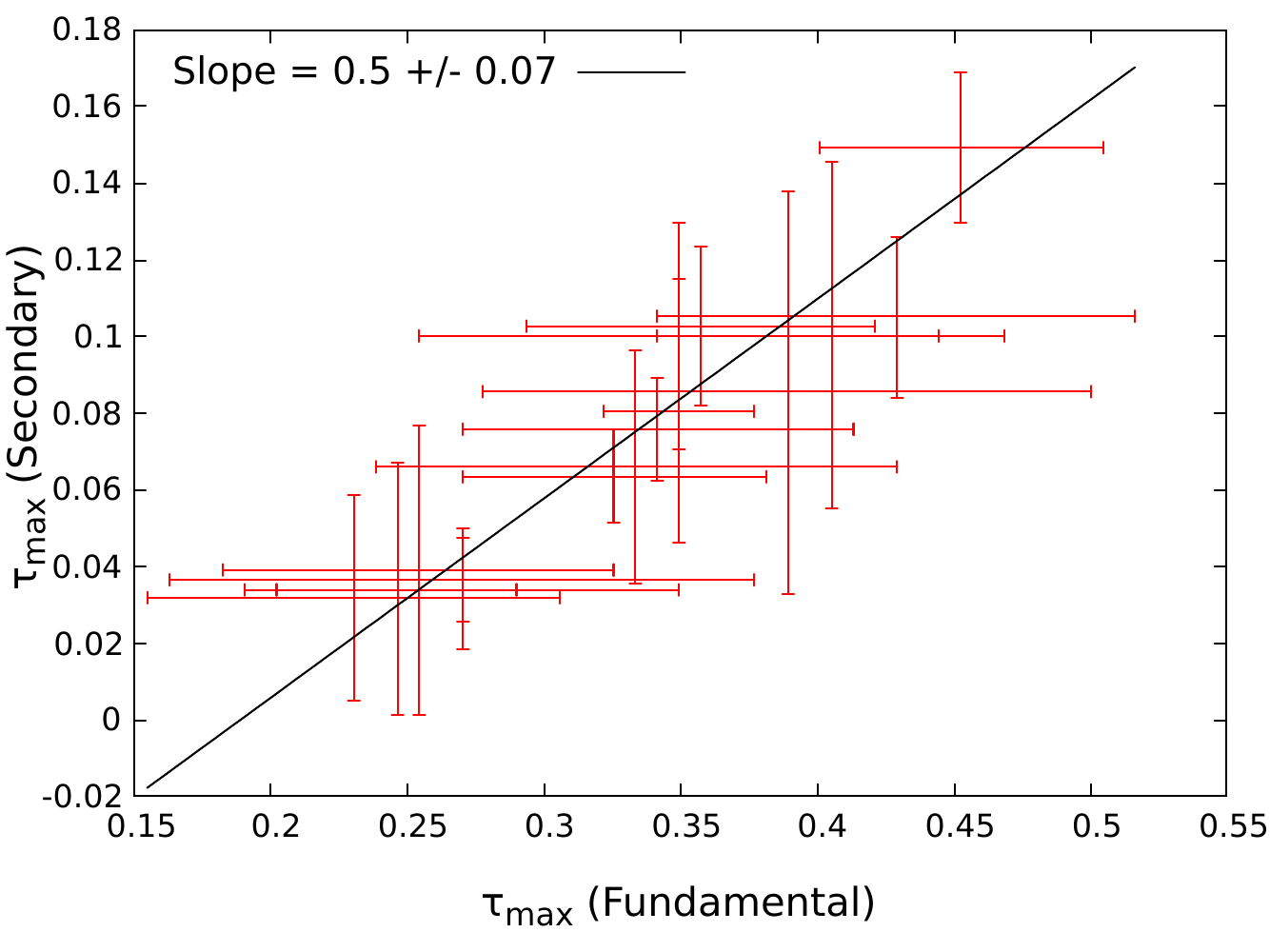}
    \end{subfigure}
    \begin{subfigure}[b]{0.35\textwidth}
        \centering
        \includegraphics[trim=0.0cm 0.0cm 0.0cm 0.0cm, clip, width=\textwidth]{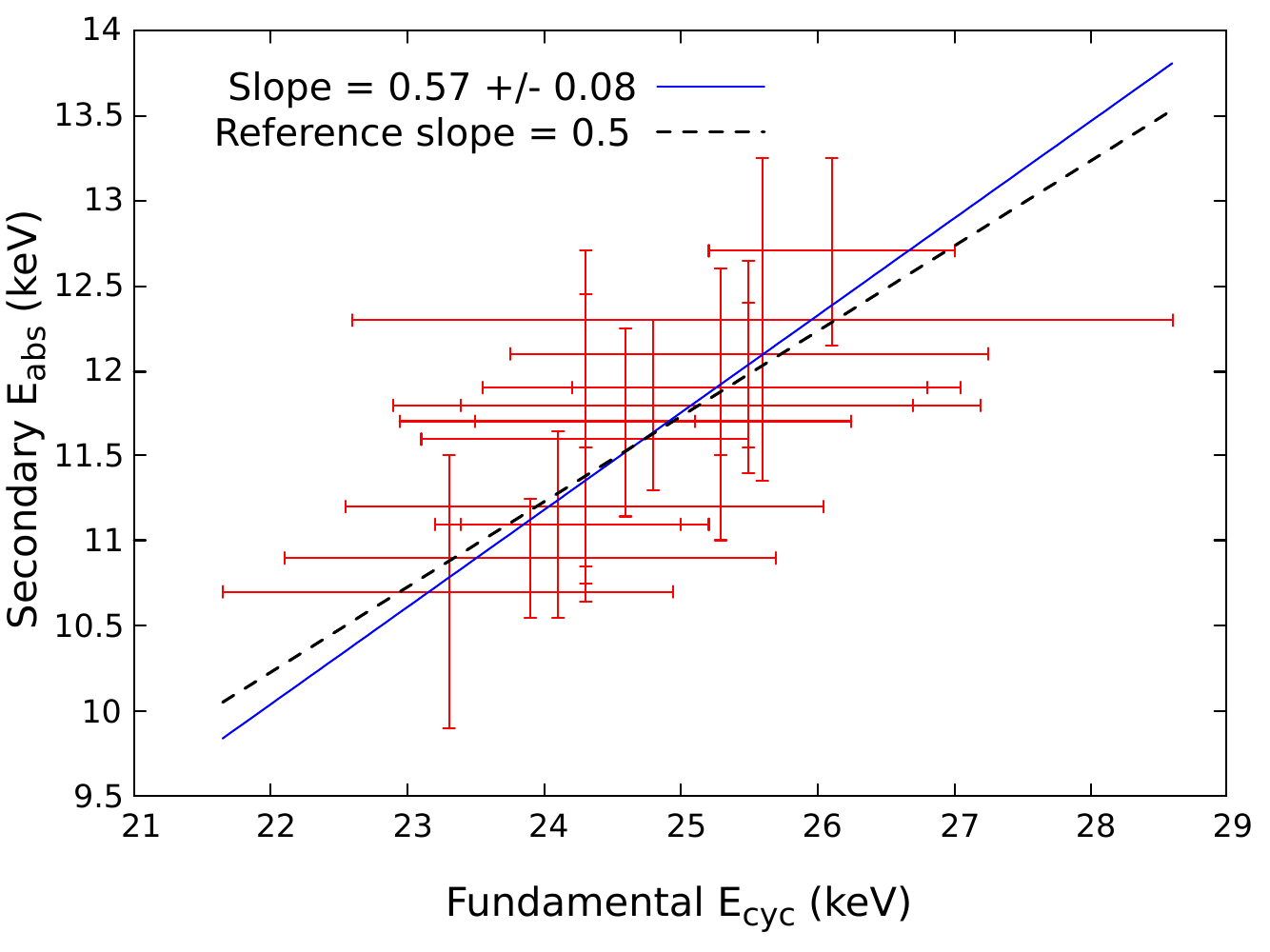}
    \end{subfigure}
    \caption{For ASTROSAT data, Correlation of secondary absorption feature with fundamental CRSF feature. \textbf{Top}: Secondary absorption optical depth vs fundamental CRSF optical depth. A linear fit gives, slope m = 0.5  $\pm$  0.07 and y-intercept c = -0.1. Slope suggests secondary absorption depth to be $\sim$ 50\% of fundamental. \textbf{Bottom}: Secondary absorption energy vs fundamental CRSF absorption energy. Slope of 0.57 $\pm$ 0.08 gives the ratio of energies, dashed line is the reference line with slope 0.5 for weak energy to be exactly half of fundamental. Observed slope falls within 1$\sigma$ deviation from the reference line.}
    \label{fig:astroweakfund}
\end{figure}

Thus, in \texttt{XSPEC}, we implemented the continuum model \texttt{Tbabs*(pcfabs*gabs*gabs*newhcut*powerlaw+gaussian)} for final analysis. Width of fundamental CRSF was close to 5 keV and secondary was close to 1.6 keV. We fit the widths with errors with a constant function and got $\chi^{2}_{red}$ $\sim$ 1 for value 5.02 $\pm$ 0.068 keV for fundamental line width and 1.634 $\pm$ 0.029 keV for secondary absorption line width.
\begin{table}
\small
\setlength{\tabcolsep}{4.5pt}
    \centering
    \caption{Few parameters in the spectral fit with the continuum model for ASTROSAT observation showed consistency over the orbital phase. We fit these parameters with errors with a constant function \texttt{f(x) = constant}. The best fit values with errors are provided below.}
    \begin{tabular}{@{}ccccc@{}}
        \toprule
        $NH_{1}^a$& $\sigma_{weak}^b$ & $\sigma_{fund}^b$ & $E_{cutoff}^b$ & $Energy_{K_{\alpha}}^b $ \\
        \midrule
        &&&&\\
        $0.63\pm 0.01$ & $1.63\pm0.03$ & $5.02\pm0.07$ & $14.02\pm0.04$ & $6.43\pm0.01$ \\
        &&&&\\
       \bottomrule
    \end{tabular}
    \RaggedRight \\ \textbf{Notes: }${}^a$ Additional hydrogen column density in units $10^{22}$ \linebreak ${}^b$ In units of keV.
    \label{tab:fixed_params}
\end{table}
Cutoff energy also did not show variations over the orbital phase and was close to 14 keV and hence was fixed. Details of parameters, that were fixed for the final fitting are provided in Table ~\ref{tab:fixed_params}. Best fit spectra for two reference segments, one in high variability state (segment 4) and other in uniform intensity state (segment 13) is shown in Figure ~\ref{fig:astrospec}. The best fit parameter values with errors for time resolved analysis are given in Table~\ref{tab:Table2}.
\begin{figure*}
    \resizebox{18cm}{6cm}{\includegraphics[width=\textwidth]{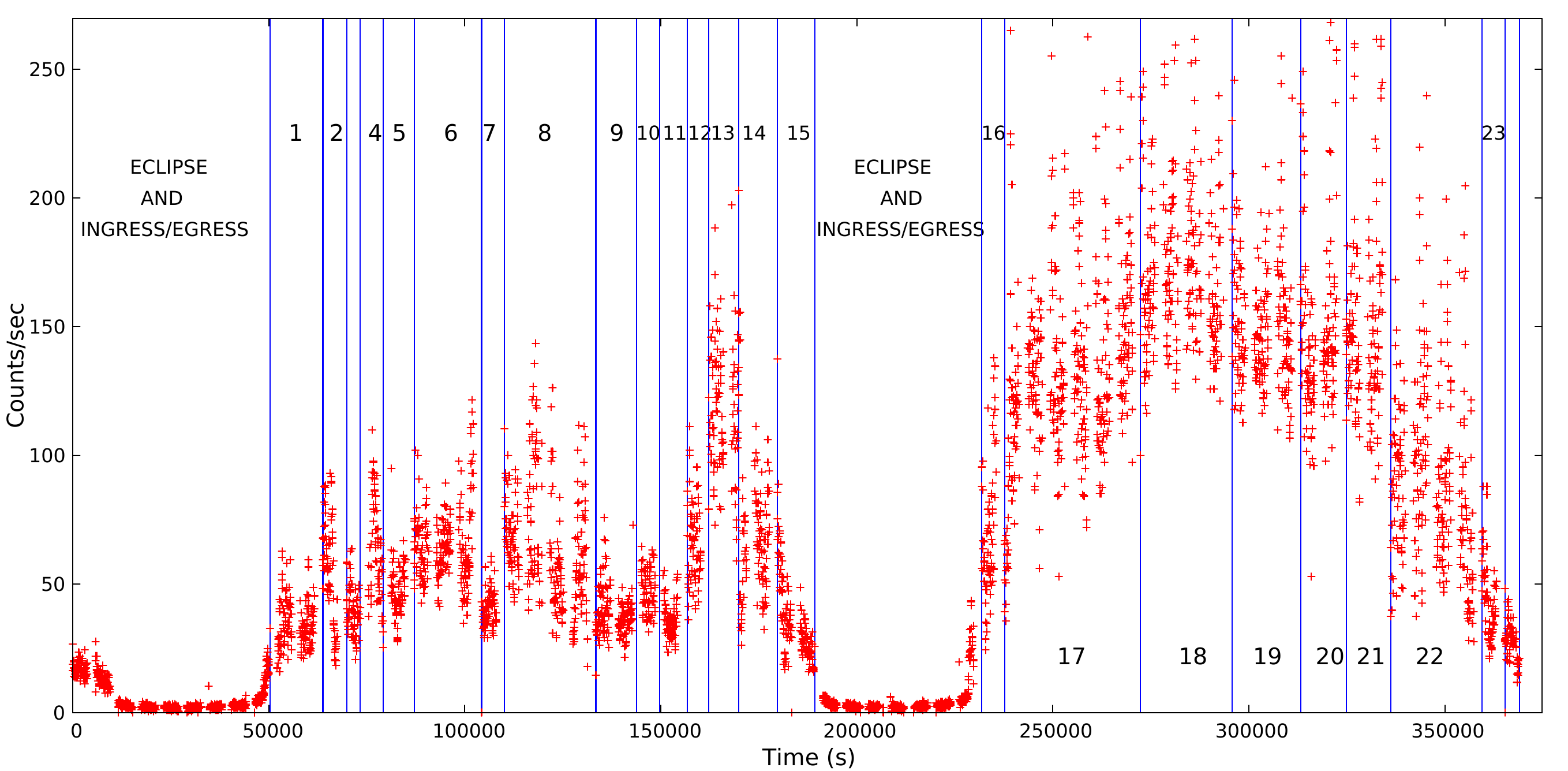}}
    \caption{Segmentation of NuSTAR FPMA light curve for Cen X-3. Segments 1 to 15 are in first binary orbit in low flux state in non-eclipse region. Segments 7, 9 and 11 are dips in low state. Segment 13 is a bump in low state. Segments 16 to 23 are in high flux state in second binary orbit. Segment 24 (not marked) is the eclipse ingress post second orbit. \\\\} 
    \label{fig:nu_segments}
\end{figure*}
Iron line emission flux is $\sim$ 0.4 $\times 10^{-10}$ ergs  cm$^{-2}$ s$^{-1}$ with a slight variation of equivalent width from 0.02 to 0.05 keV over orbital phase. We did not see correlation of secondary absorption energy with cutoff energy. Optical depth is calculated from strength by the following relation :
\begin{equation}
\label{eq:eq2}
   \hspace{1.5cm} \text{Optical}\,\text{depth}\,(\tau_{\text{max}}) = \left(\frac{\text{Strength}}{\text{width} \times \sqrt{2\pi} } \right)\\
\end{equation}
Optical depth as well as energy of fundamental CRSF absorption showed a non-linear anti-correlation with 1 to 30 keV flux (Figure ~\ref{fig:astrofund}) described by a powerlaw:
\begin{equation}
\label{eq:eq1}
   \hspace{1.5cm} A = A_0 \left(\frac{f}{10^{-10}\; \text{ergs}\,\text{cm}^{-2}\,\text{s}^{-1}} \right)^{-\beta}\\
\end{equation}
where \textbf{A} represents a function that can be either energy or optical depth. $A_0$ is energy normalisation in units of keV for \textbf{A} representing cyclotron line energy. $A_0$ is a dimensionless optical depth normalisation for \textbf{A} representing optical depth. $\beta$ is the anti-correlation index and \textbf{f} is the flux in units of $10^{-10}$ ergs  cm$^{-2}$ s$^{-1}$. Secondary absorption showed similar trend as that of fundamental CRSF absorption (Figure ~\ref{fig:astroweak}). Both fundamental and secondary absorption energy vs flux was well fitted with $\beta$ of 0.08 $\pm$ 0.01 with variation in normalisation. Depth of fundamental have $\beta$ of 0.4 $\pm$ 0.01 and secondary have $\beta$ of 1.1 $\pm$ 0.2. Fundamental CRSF absorption and secondary absorption showed positive correlation with each other in regards to both depth and energy (Figure ~\ref{fig:astroweakfund}). Optical depth of this absorption varied within 0.02 and 0.15. Secondary absorption and fundamental CRSF absorption depth were related by ratio of 0.5  $\pm$  0.07 ($\sim$ 50\% of fundamental). Secondary absorption energy and fundamental CRSF energy were related by ratio of 0.57  $\pm$  0.08. 

\subsection{NuSTAR}
\vspace{5pt}
Segmentation of NuSTAR FPMA light curve for Cen X-3 is shown in Figure ~\ref{fig:nu_segments}. NuSTAR covered almost two binary orbit and light curve showed count rate variation by a factor of $\sim$ 3 in second orbital phase. We have excluded the eclipses, ingress and egress regions for the present work. Initially we adopted the best fit model of ASTROSAT to 3 to 30 keV NuSTAR data for full exposure. Model parameters were kept free initially, which produced fit with $\chi^{2}_{red}$ $\sim$ 1.6. From SUZAKU observation of 2008, XIS has clearly detected three iron emission lines in X-ray spectrum of Cen X-3 at 6.4, 6.7 and 6.9 keV \citep{Naik} and also previously by ASCA \citep{ebis96}. In addition to 6.4 keV line we noted emission $\sim$ 6.9 keV and $\sim$ 5.7 keV. 6.7 keV line was not seen. Similar to ASTROSAT result, a weak secondary absorption feature was also seen but at slightly higher value in energy. Cutoff component \texttt{highecut} and \texttt{newhcut} did not show much variations in the parameters, except the strength of the weak absorption, which was found to be comparatively higher with \texttt{highecut}. We adopted \texttt{newhcut} as our cutoff component for analysis and applied this model (\texttt{tbabs*(pcfabs*gabs*gabs*newhcut*po+ga+ga+ga)}) to the time resolved segments.

\begin{figure*}
    \centering
    \begin{subfigure}[b]{0.45\textwidth}
         \centering
         \includegraphics[width=\textwidth]{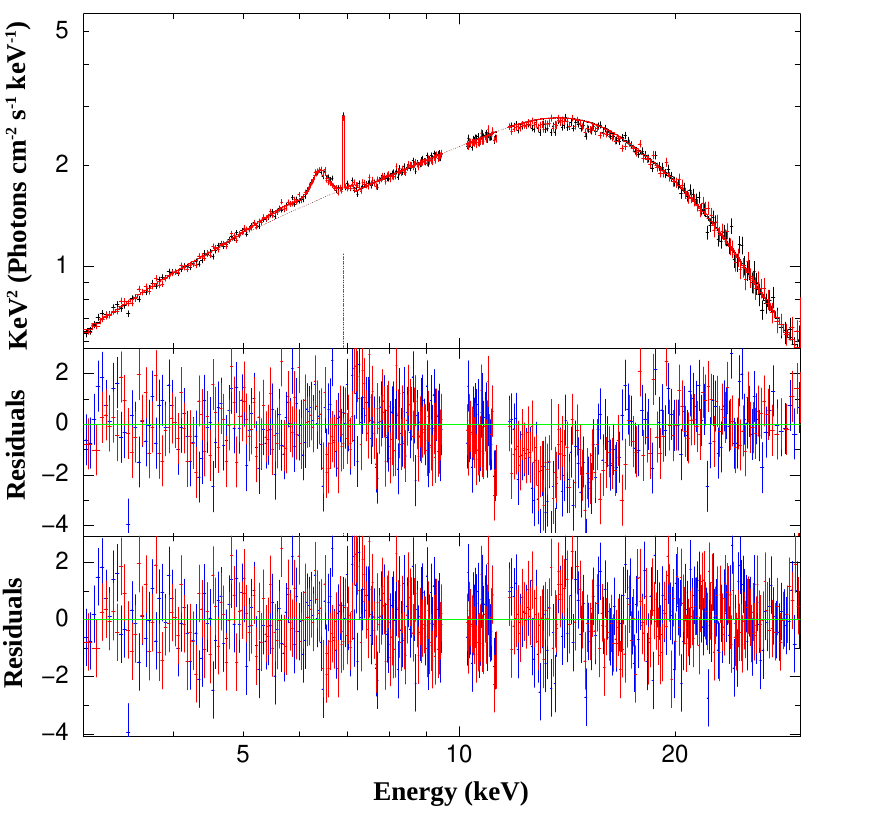}
         \caption{}
    \end{subfigure}
    \begin{subfigure}[b]{0.45\textwidth}
         \centering
         \includegraphics[width=\textwidth]{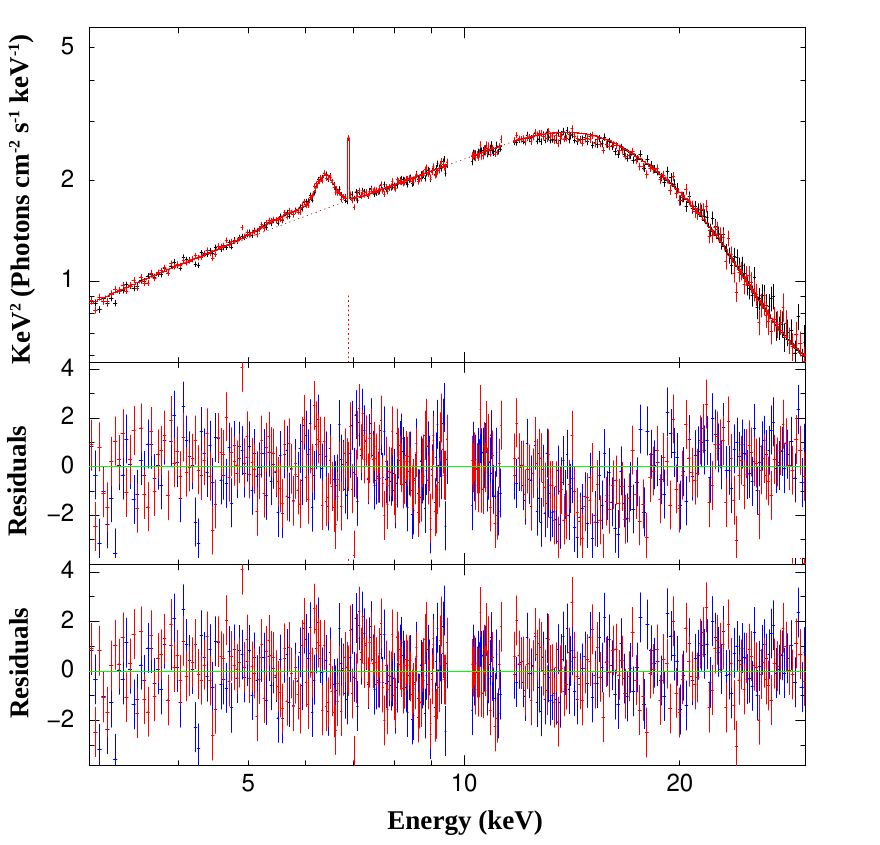}
         \caption{}
    \end{subfigure}
    \caption{\textbf{a}: Best fit spectra from joint fit of FPMA and FPMB observation onboard NuSTAR for segment 8 in first binary orbit. \textbf{Upper panel}: Best fit unfolded spectra. \textbf{Middle panel}: Best fit residuals for phenomenological model without secondary absorption component. \textbf{Bottom panel}: Best fit residuals for phenomenological model with secondary absorption component. \textbf{b}: Similar as (a) for segment 18 in second binary orbit. Segments 8 and 18 are from Figure ~\ref{fig:nu_segments}.\\\\\\}
    \label{fig:nuspec}
\end{figure*}
\begin{figure*}
    \centering
    \begin{subfigure}[b]{0.36\textwidth}
        \centering
            \includegraphics[trim=0.0cm 0.0cm 0.0cm 0.0cm, clip, width=\textwidth]{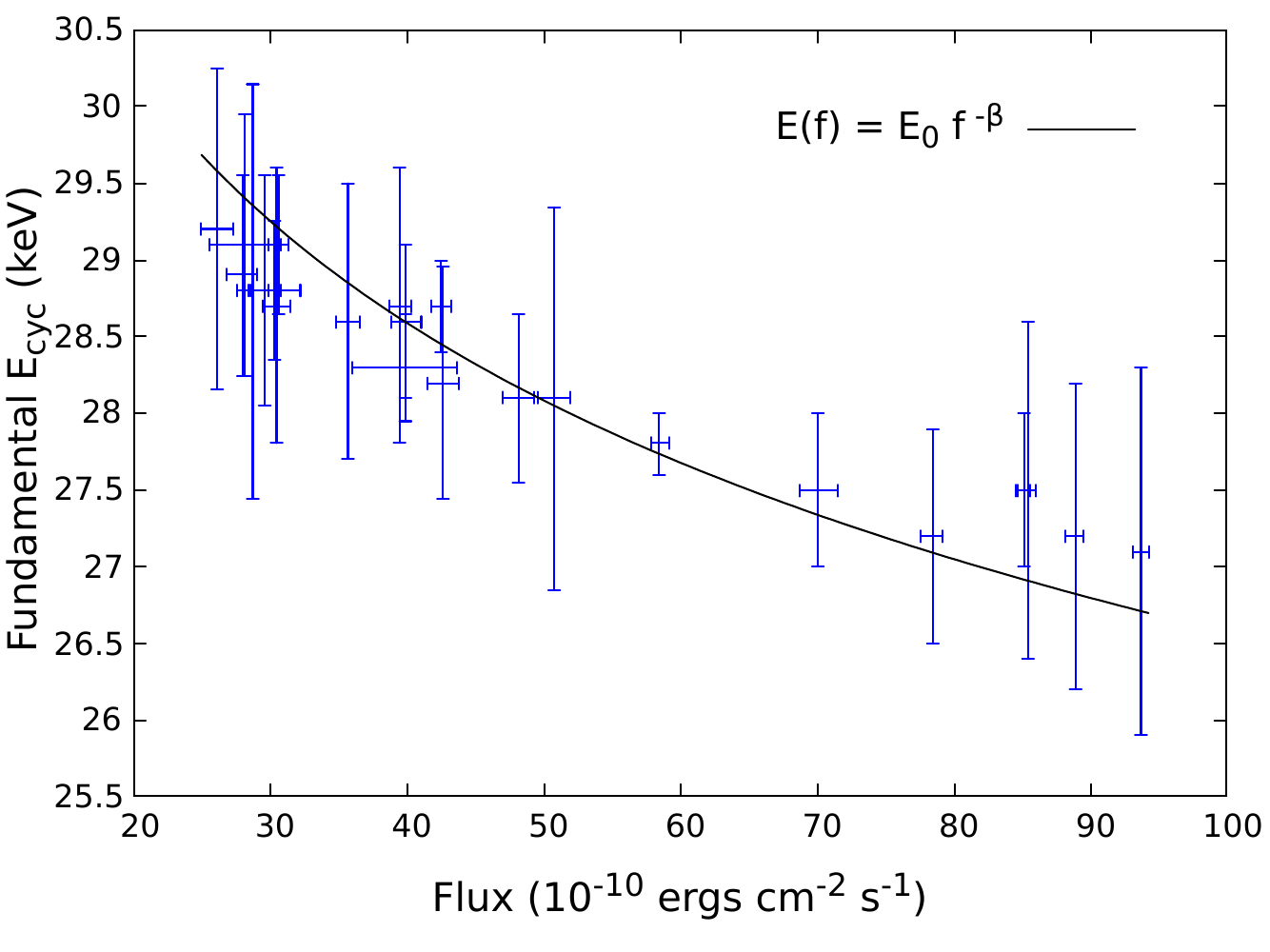} 
            \caption{}
    \end{subfigure}
    \begin{subfigure}[b]{0.36\textwidth}
        \centering
        \includegraphics[trim=0.0cm 0.0cm 0.0cm 0.0cm, clip, width=\textwidth]{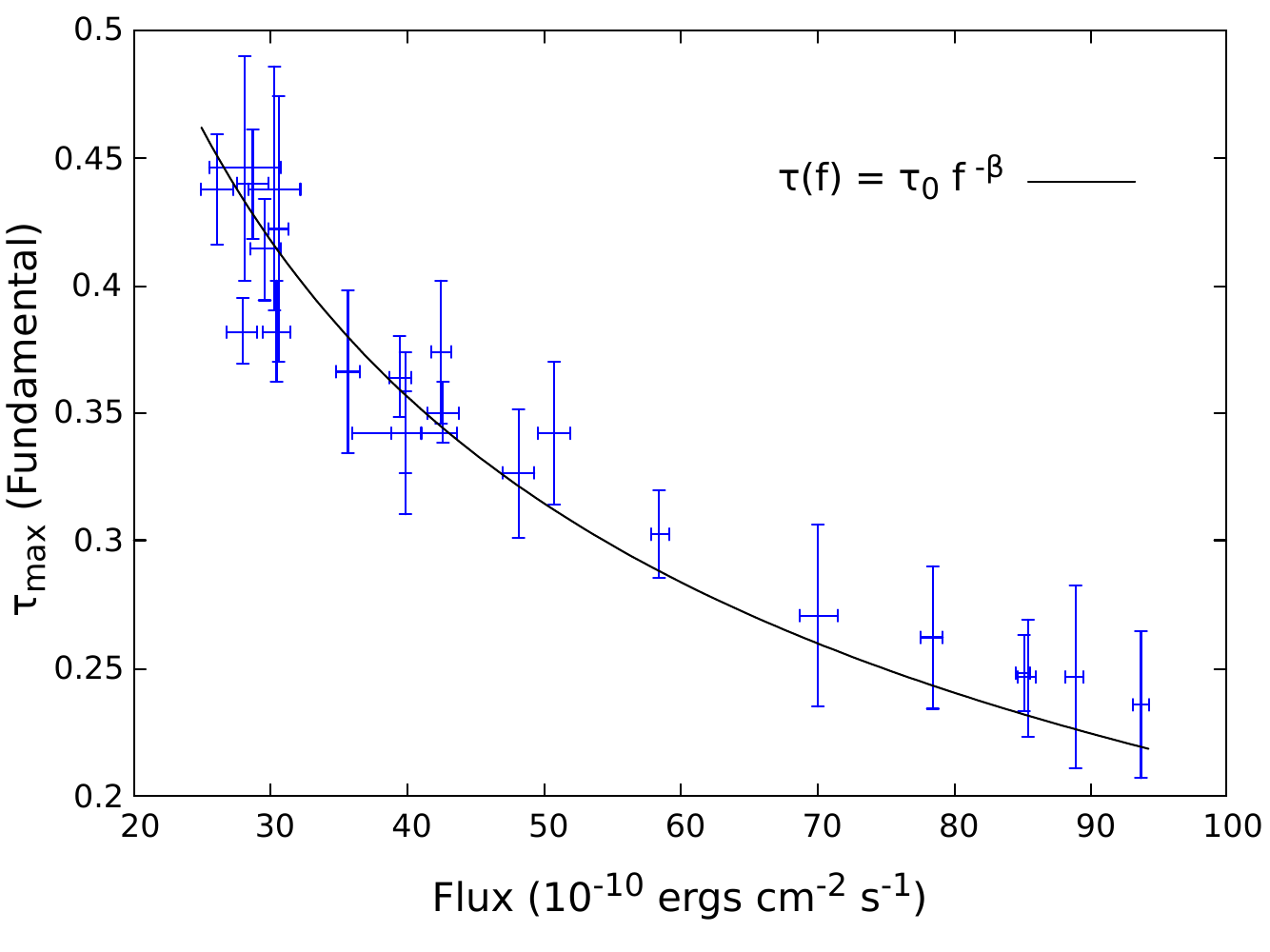}
        \caption{}
        \label{fig:nucorr2}
    \end{subfigure}
    \caption{For NuSTAR data, \textbf{a}: Fundamental CRSF absorption energy vs 3 to 30 keV flux showing powerlaw anti-correlation trend with index $\beta$ = 0.08 $\pm$ 0.02 and normalisation E$_0$ = 38.3 $\pm$ 0.08. \textbf{b}: Powerlaw anti-correlation of index $\beta$ = 0.5 $\pm$ 0.003 and normalisation $\tau_0$ = 2.5 $\pm$ 0.03 between fundamental CRSF optical depth and flux.\\\\\\}
    \label{fig:nufund}
\end{figure*}
\begin{figure*}
    \centering
    \begin{subfigure}[b]{0.36\textwidth}
        \centering
            \includegraphics[trim=0.0cm 0.0cm 0.0cm 0.0cm, clip, width=\textwidth]{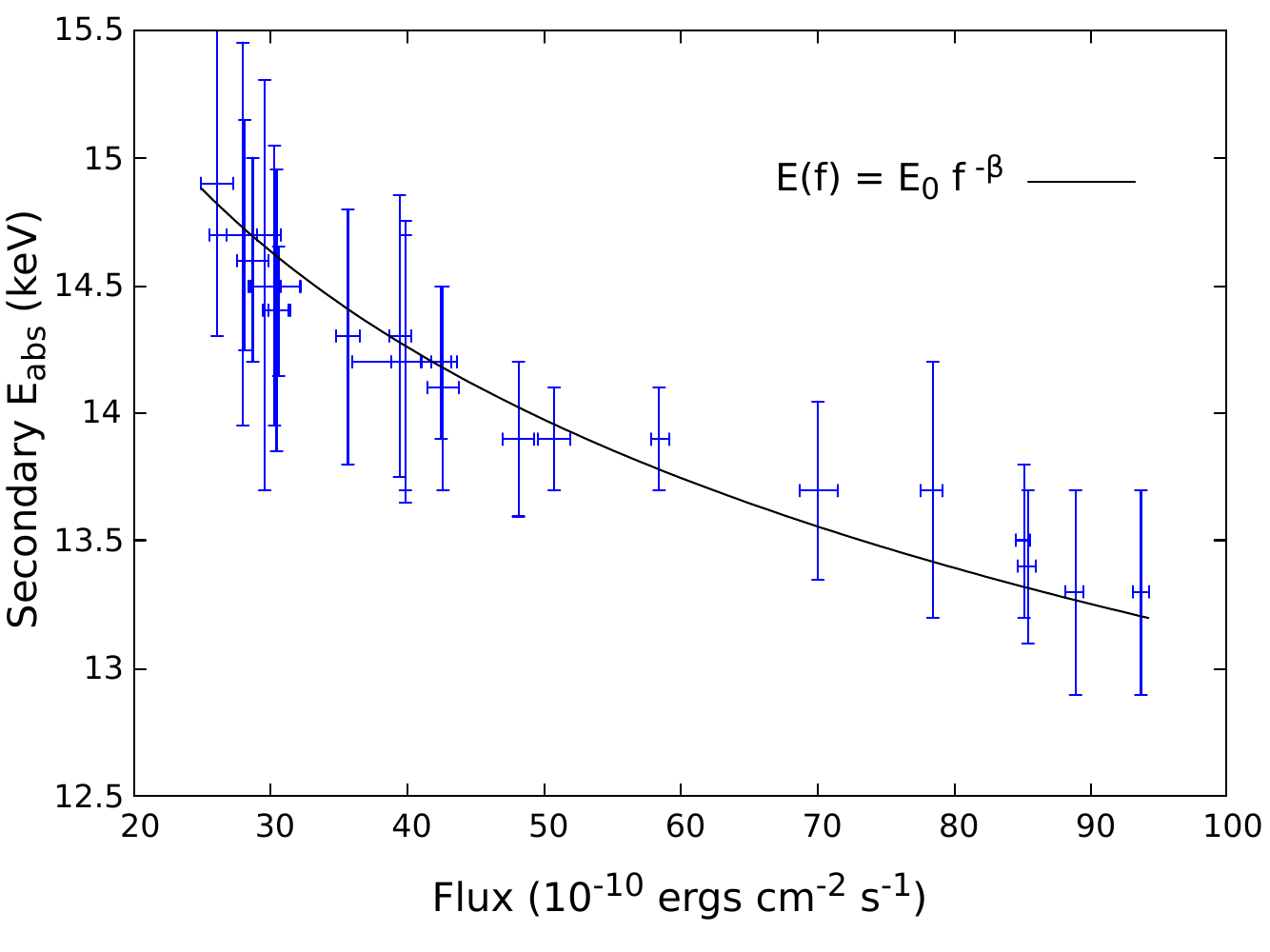} 
            \caption{}
    \end{subfigure}
    \begin{subfigure}[b]{0.36\textwidth}
        \centering
        \includegraphics[trim=0.0cm 0.0cm 0.0cm 0.0cm, clip, width=\textwidth]{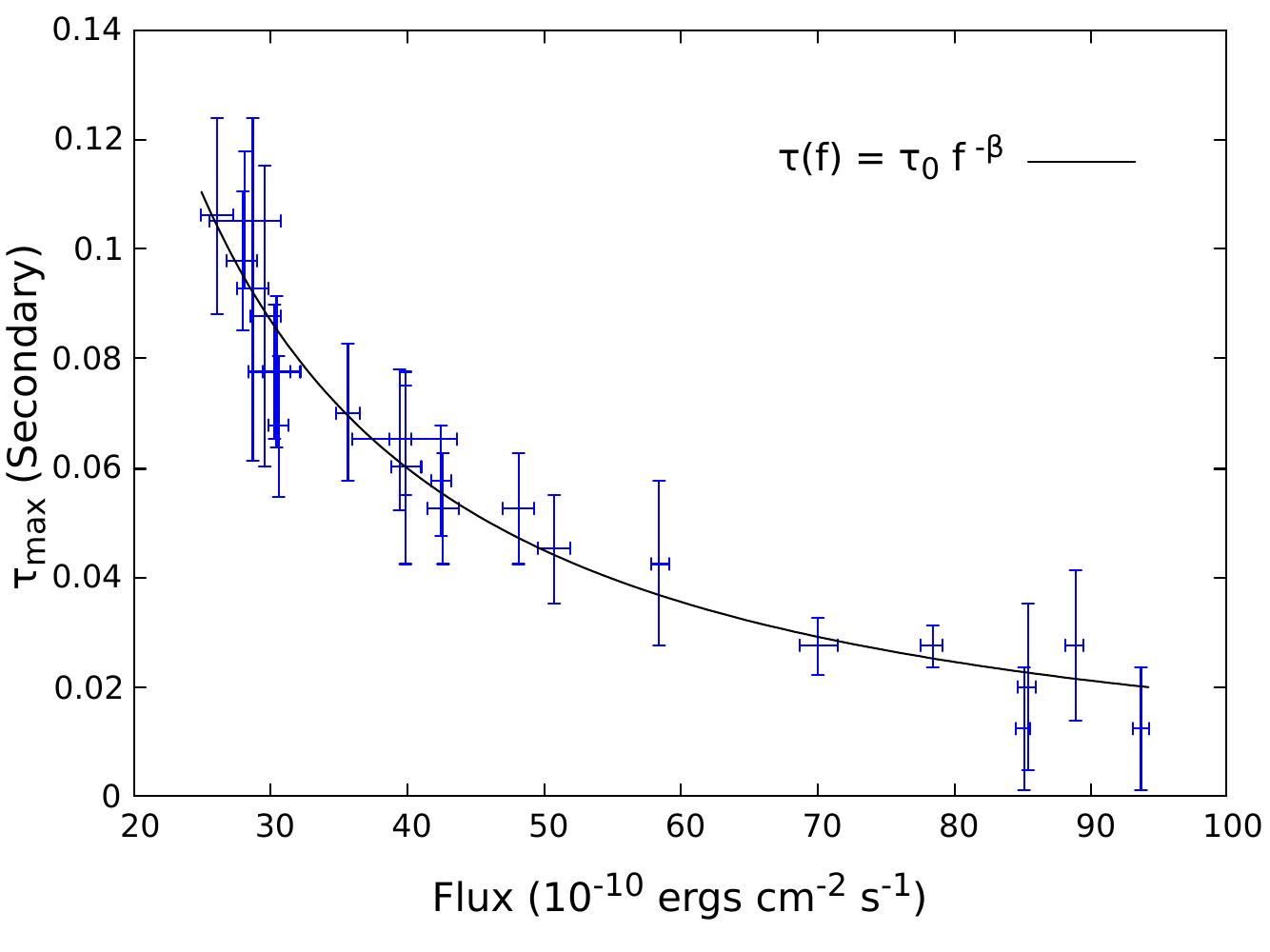}
        \caption{}
    \end{subfigure}
    \caption{For NuSTAR data, \textbf{a}: Secondary absorption energy vs 3 to 30 keV flux showing powerlaw anti-correlation trend with index $\beta$ = 0.09 $\pm$ 0.01 and normalisation E$_0$ = 19.9 $\pm$ 0.04. \textbf{b}: Powerlaw anti-correlation of index $\beta$ = 1.3 $\pm$ 0.02 and normalisation $\tau_0$ = 7.2 $\pm$ 0.2 between secondary absorption optical depth and flux.\\\\\\}
    \label{fig:nuweak}
\end{figure*}
\begin{figure*}
    \centering
    \begin{subfigure}[b]{0.36\textwidth}
        \centering
            \includegraphics[trim=0.0cm 0.0cm 0.0cm 0.0cm, clip, width=\textwidth]{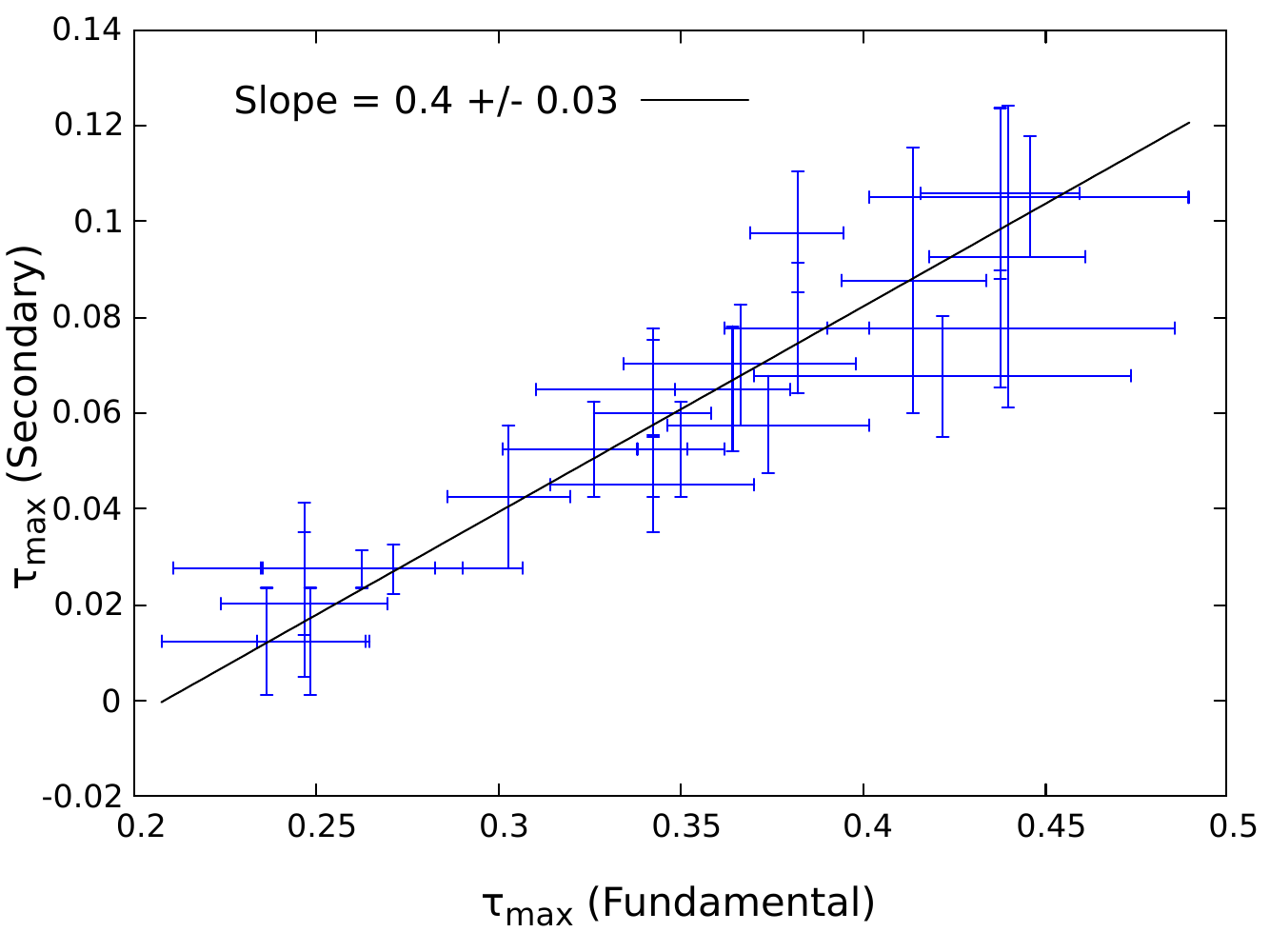}
            \caption{}
    \end{subfigure}
    \begin{subfigure}[b]{0.36\textwidth}
        \centering
        \includegraphics[trim=0.0cm 0.0cm 0.0cm 0.0cm, clip, width=\textwidth]{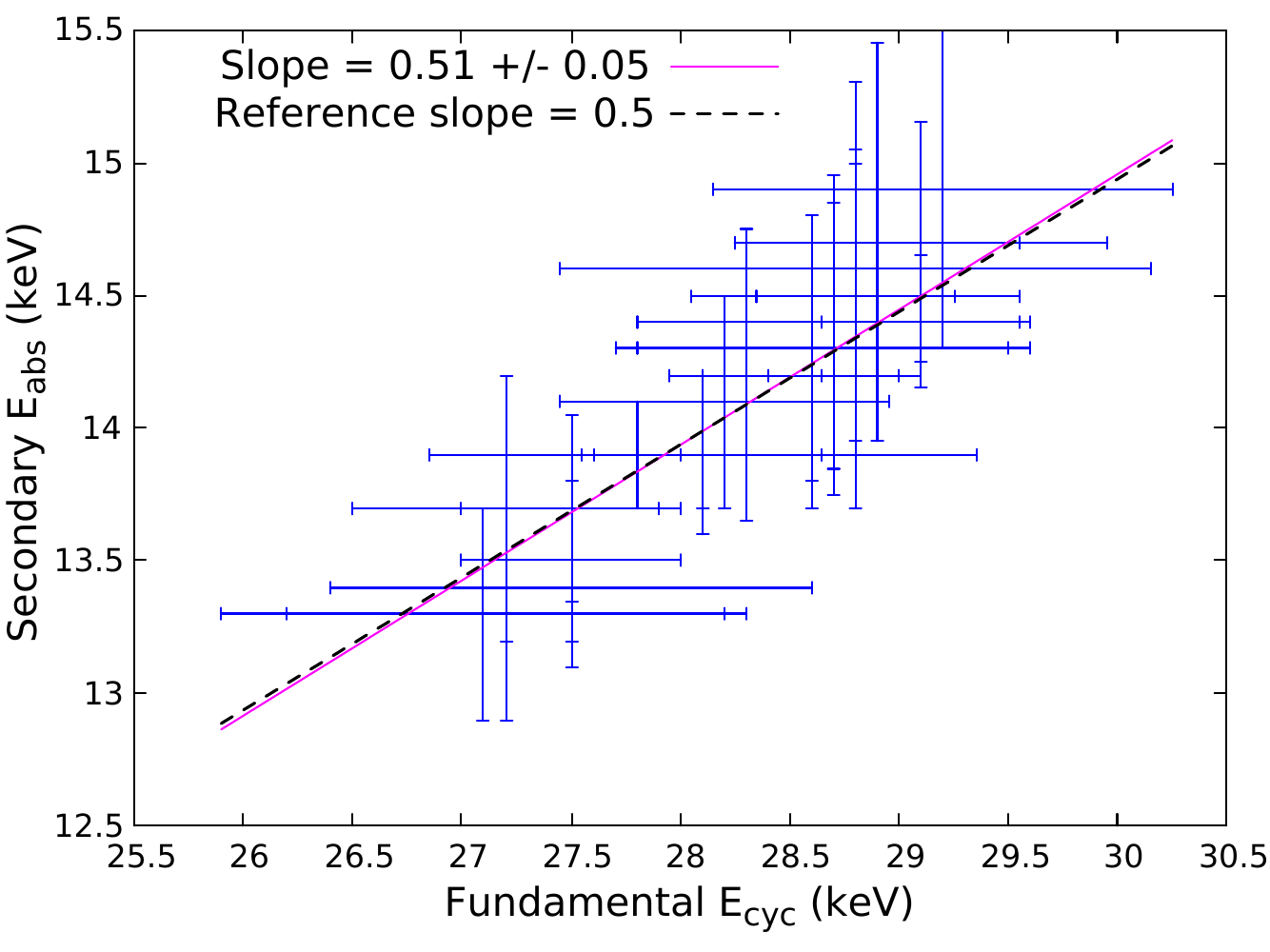}
        \caption{}
    \end{subfigure}
    \caption{\small For NuSTAR data, Correlation of secondary absorption feature with fundamental CRSF feature. \textbf{a}: Secondary absorption optical depth showing a positive correlation to fundamental CRSF optical depth. Slope of 0.4  $\pm$  0.03 suggests secondary absorption depth to be $\sim$ 40\% of fundamental. \textbf{b}: Secondary absorption energy vs fundamental CRSF absorption energy showing positive correlation. Slope of 0.51 $\pm$ 0.05 gives the ratio of energies, dashed line is the reference line with slope 0.5 for weak energy to be exactly half of fundamental. Observed slope falls within 1$\sigma$ deviation from the reference line.}
    \label{fig:nuweakfund}
\end{figure*}
We noticed two consecutive sharp absorption like feature in NuSTAR spectrum in addition to above features from 9.5 to 10.2 keV and 11.3 to 11.7 keV. We were unsure whether these features were real or artifact. Including these features did not result in a substantial alteration of the best fit parameter values, albeit they marginally elevated the chi-squared value for the overall fit. We excluded this energy range in the spectral analysis of all the segments. Hydrogen column density for interstellar absorption (\textit{nH}1) was close to that obtained from ASTROSAT data for all segments in the first binary orbit excepting segment 13. \textit{nH}1 could not be constrained in the second binary orbital phase which is in high flux state, so we fixed this value to 0.64 $\times 10^{22}$ units. Cutoff energy was close to 14 keV in all the segments. The secondary absorption energy varied within 13.5 to 15.3 keV. Similar to ASTROSAT results, secondary absorption energy showed no correlation with the powerlaw cutoff energy and its width was close to 1.6 keV. Energy of the fundamental cyclotron line showed variation between 27.1 to 29.9 keV over orbital phase. Fundamental CRSF line width was close to 5 keV. Iron K$_{\alpha}$ line emission energies were close to 5.7, 6.4 and 6.9 keV. We have fitted these parameters with a constant function and fixed their values. Details of parameters are provided in Table ~\ref{tab:nu_fixed_params}. Best fit spectra for two reference segments, one in first binary orbit in low flux state (segment 8) and other in second binary orbit in high flux state (segment 18) are shown in Figure ~\ref{fig:nuspec}. The best fit parameter values with errors for time resolved analysis are given in Table~\ref{tab:nu_fit}.

\begin{table}
\small
\setlength{\tabcolsep}{28pt}
\renewcommand{\arraystretch}{1.5}
    \centering
    \caption{Few parameters in the spectral fit with the continuum model for NuSTAR observation showed consistency over the orbital phase. We fit these parameters with errors with a constant function \texttt{f(x) = constant} and fixed them for spectral analysis. The best fit values with errors are provided below.}
    \begin{tabular}{cc}
        \toprule
        \textbf{Parameter} & \textbf{Fit value with error}\\
        \midrule
        $NH_{1}^a$& 0.64 $\pm$ 0.001\\
        $\sigma_{weak}^b$ & 1.59 $\pm$ 0.003\\
        $\sigma_{fund}^b$ & 5.01 $\pm$ 0.02\\
        $E_{cutoff}^b$ & 13.95  $\pm$ 0.04 \\
        $Energy_{K_{\alpha}}^b $ & 5.69 $\pm$ 0.01\\
        $Energy_{K_{\alpha}}^b $& 6.42 $\pm$ 0.01\\
        $Energy_{K_{\alpha}}^b $ &6.87 $\pm$  0.01\\
        \bottomrule
        \end{tabular}
    \RaggedRight \\ \textbf{Notes: }${}^a$ Additional hydrogen column density in units $10^{22}$ \linebreak ${}^b$ In units of keV.
    \label{tab:nu_fixed_params}
\end{table}
5.7 keV emission line flux varied within 0.02 and 0.35 $\times$ 10$^{-10}$ ergs\,cm$^{-2}$\,s$^{-1}$ and equivalent width varied within 0.04 to 0.19 keV. 6.4 keV line flux varied within 0.16 and 0.46 $\times$ 10$^{-10}$ ergs\,cm$^{-2}$\,s$^{-1}$ and equivalent width within 0.1 and 0.4 keV. 6.9 keV line flux varied within 0.01 and 0.09 $\times$ 10$^{-10}$ ergs\,cm$^{-2}$\,s$^{-1}$ and equivalent width within 0.01 and 0.08 keV. 

Correlations of CRSF absorption feature and secondary absorption feature with 3 to 30 keV flux and between each other (in regards to depth and energies) were similar to ASTROSAT. Strength as well as energy of fundamental CRSF absorption showed a non linear anti-correlation with the flux (Figure ~\ref{fig:nufund}) given by equation ~\ref{eq:eq1}. Similar trend was shown by the secondary absorption strength and energy to flux (Figure ~\ref{fig:nuweak}). Similar to ASTROSAT, fundamental and secondary absorption energy vs flux was well fitted with $\beta$ of 0.08 $\pm$ 0.01 and 0.09 $\pm$ 0.01 respectively. Depth of fundamental have $\beta$ of 0.5 $\pm$ 0.003 and secondary have $\beta$ of 1.3 $\pm$ 0.02.

Secondary and fundamental absorption strength as well as energy showed positive correlation to each other. Secondary absorption and fundamental CRSF absorption was found to have depth ratio of 0.4 $\pm$ 0.03 ($\sim$ 40 \% of fundamental) and energy ratio of 0.51 $\pm$ 0.05 (Figure ~\ref{fig:nuweakfund}). It is seen that the slope lies within 1$\sigma$ deviation from the reference slope (Bottom panel of Figure ~\ref{fig:nuweakfund}). The presence of weak and narrow absorption line in the vicinity of $\sim$ 14.4 keV with width $\sim$ 1.6 keV is seen throughout the segments similar to ASTROSAT.

\begin{table*}
\centering
\small
\setlength{\tabcolsep}{3.2pt} 
\renewcommand{\arraystretch}{1.7} 
\caption{Best fit parameters with ASTROSAT data for Cen X-3. Model: tbabs*(pcfabs*gabs*gabs*newhcut*powerlaw + gaussian). Hydrogen column density for Interstellar absorption is fixed at $0.625 \times 10^{22}$ units, width of secondary absorption fixed at 1.634 keV and  of fundamental at 5 keV, cutoff energy is fixed at 14.08 keV and smoothing width at 5 keV. Errors are within 90 \% confidence.}
\label{tab:Table2}
\begin{tabular}{cccccccccccccc} 
 \toprule
    \textbf{Sgmt} &
      \multicolumn{2}{c}{\textbf{pcfabs}}& \multicolumn{1}{c}{\textbf{Unabs}} &\multicolumn{2}{c}{\textbf{gabs {weak}}} & \multicolumn{2}{c}{\textbf{gabs {fund}}} &\multicolumn{1}{c}{\textbf{newhcut}} & \multicolumn{2}{c}{\textbf{powerlaw}} & \multicolumn{2}{c}{\textbf{6.43 keV line}} &\multicolumn{1}{c}{\textbf{$\chi^{2}_{red}$}}\\
      \cmidrule(l){2-3}\cmidrule(r){5-6}\cmidrule(l){7-8}\cmidrule(l){10-11}\cmidrule(l){12-13}\cmidrule(l){14-14}
       \textbf{No.} & $NH^a$ & $Frac_{cov}$ &\textbf{$Flux^b$} & $E0^c$ & $strength^c$ & $E0^c$ & $strength^c$ & $E_{fold}^{c}$  & $\Gamma$ & $norm$ & $Flux^b$ & $Eqw^c$ & $d.o.f$\\ 
\bottomrule
&&&&&&&&&&&&\\
1&$13.4_{-2.8}^{+2.8}$&$0.8_{-0.26}^{+0.14}$&$19.2_{-2.1}^{+1.9}$&$12.1_{-0.5}^{+0.6}$&$0.41_{-0.19}^{+0.18}$&$25.5_{-2.4}^{+1.1}$&$5.1_{-0.8}^{+0.8}$&$8.5_{-0.5}^{+1.1}$&$0.8_{-0.09}^{+0.08}$&$0.04_{-0.01}^{+0.01}$&$0.4_{-0.1}^{+0.1}$&$0.3_{-0.1}^{+0.1}$ & 1.01/106 \\\\
2&$11.3_{-2.1}^{+1.6}$&$0.8_{-0.06}^{+0.04}$&$32.5_{-2.4}^{+3.6}$&$11.7_{-0.5}^{+0.6}$&$0.31_{-0.06}^{+0.05}$&$24.6_{-2.4}^{+0.9}$&$4.3_{-0.4}^{+1.4}$&$9.9_{-0.6}^{+0.8}$&$0.9_{-0.05}^{+0.05}$&$0.07_{-0.01}^{+0.01}$&$0.4_{-0.1}^{+0.1}$&$0.5_{-0.1}^{+0.1}$ & 1.03/232 \\\\
3&$10.3_{-2.6}^{+2.9}$&$0.7_{-0.12}^{+0.08}$&$74.7_{-4.6}^{+5.1}$&$11.1_{-0.7}^{+0.4}$&$0.16_{-0.15}^{+0.16}$&$24.1_{-0.9}^{+0.9}$&$3.2_{-0.9}^{+0.9}$&$9.3_{-0.5}^{+0.7}$&$0.9_{-0.06}^{+0.06}$&$0.07_{-0.01}^{+0.01}$&$0.4_{-0.1}^{+0.1}$&$0.4_{-0.1}^{+0.1}$ & 1.02/191 \\\\
4&$11.3_{-2.1}^{+1.9}$&$0.8_{-0.02}^{+0.02}$&$36.9_{-2.5}^{+2.9}$&$11.7_{-0.9}^{+1.1}$&$0.27_{-0.07}^{+0.18}$&$24.3_{-0.9}^{+0.7}$&$4.2_{-0.7}^{+1.7}$&$8.2_{-0.7}^{+0.8}$&$0.9_{-0.05}^{+0.05}$&$0.08_{-0.01}^{+0.01}$&$0.4_{-0.1}^{+0.1}$&$0.5_{-0.1}^{+0.1}$ & 1.03/235 \\\\
5&$13.1_{-2.5}^{+2.6}$&$0.8_{-0.07}^{+0.04}$&$40.5_{-1.8}^{+1.8}$&$11.6_{-0.8}^{+0.9}$&$0.26_{-0.03}^{+0.07}$&$24.3_{-1.2}^{+1.2}$&$4.1_{-0.5}^{+0.9}$&$9.8_{-1.1}^{+1.4}$&$0.9_{-0.05}^{+0.05}$&$0.09_{-0.01}^{+0.01}$&$0.4_{-0.1}^{+0.1}$&$0.5_{-0.1}^{+0.1}$ & 1.12/137 \\\\
6&$10.6_{-1.6}^{+1.1}$&$0.7_{-0.09}^{+0.06}$&$76.3_{-4.4}^{+4.8}$&$10.9_{-0.4}^{+0.3}$&$0.13_{-0.01}^{+0.21}$&$23.9_{-2.7}^{+0.9}$&$2.9_{-1.1}^{+0.8}$&$8.8_{-1.9}^{+3.6}$&$0.9_{-0.05}^{+0.09}$&$0.08_{-0.01}^{+0.01}$&$0.3_{-0.1}^{+0.1}$&$0.5_{-0.1}^{+0.1}$ & 1.01/265 \\\\
7&$11.2_{-1.4}^{+1.4}$&$0.8_{-0.17}^{+0.08}$&$63.5_{-4.2}^{+4.4}$&$11.1_{-0.4}^{+0.5}$&$0.14_{-0.02}^{+0.11}$&$24.3_{-0.9}^{+0.9}$&$3.4_{-0.9}^{+1.1}$&$9.3_{-3.3}^{+1.1}$&$0.9_{-0.01}^{+0.06}$&$0.07_{-0.01}^{+0.01}$&$0.4_{-0.1}^{+0.1}$&$0.4_{-0.1}^{+0.1}$ & 1.01/194 \\\\
8&$17.6_{-2.4}^{+1.6}$&$0.9_{-0.02}^{+0.02}$&$22.2_{-2.2}^{+2.8}$&$11.8_{-0.9}^{+0.7}$&$0.41_{-0.12}^{+0.12}$&$25.3_{-1.9}^{+1.9}$&$4.4_{-1.7}^{+0.7}$&$9.5_{-0.6}^{+0.8}$&$1.1_{-0.08}^{+0.08}$&$0.07_{-0.01}^{+0.01}$&$0.3_{-0.1}^{+0.1}$&$0.2_{-0.1}^{+0.1}$ & 1.05/121 \\\\
9&$10.9_{-1.9}^{+1.9}$&$0.8_{-0.18}^{+0.09}$&$86.2_{-7.4}^{+8.6}$&$10.7_{-0.8}^{+0.8}$&$0.14_{-0.13}^{+0.14}$&$23.3_{-0.6}^{+2.7}$&$3.1_{-0.2}^{+0.9}$&$8.8_{-1.1}^{+1.6}$&$0.9_{-0.03}^{+0.02}$&$0.06_{-0.02}^{+0.02}$&$0.5_{-0.1}^{+0.1}$&$0.5_{-0.1}^{+0.1}$ & 1.04/104 \\\\
10&$15.9_{-2.1}^{+2.1}$&$0.8_{-0.12}^{+0.07}$&$25.9_{-3.3}^{+5.1}$&$11.8_{-0.5}^{+0.5}$&$0.35_{-0.21}^{+0.22}$&$24.8_{-2.7}^{+1.1}$&$4.9_{-1.5}^{+1.3}$&$8.9_{-1.2}^{+1.8}$&$0.8_{-0.08}^{+0.07}$&$0.06_{-0.01}^{+0.01}$&$0.4_{-0.1}^{+0.1}$&$0.3_{-0.1}^{+0.1}$ & 1.03/272 \\\\
11&$15.7_{-2.2}^{+2.2}$&$0.8_{-0.04}^{+0.03}$&$22.4_{-1.5}^{+1.9}$&$11.8_{-1.1}^{+1.2}$&$0.27_{-0.16}^{+0.11}$&$25.3_{-1.4}^{+0.9}$&$5.5_{-0.6}^{+0.7}$&$9.8_{-0.8}^{+1.2}$&$1.1_{-0.05}^{+0.05}$&$0.07_{-0.01}^{+0.01}$&$0.4_{-0.1}^{+0.1}$&$0.4_{-0.1}^{+0.1}$ & 1.02/168 \\\\
12&$15.1_{-1.8}^{+1.9}$&$0.8_{-0.08}^{+0.05}$&$62.5_{-3.9}^{+4.2}$&$11.2_{-0.3}^{+0.4}$&$0.15_{-0.06}^{+0.03}$&$24.3_{-1.9}^{+1.6}$&$3.4_{-0.8}^{+1.9}$&$9.4_{-0.8}^{+1.1}$&$0.8_{-0.06}^{+0.05}$&$0.08_{-0.01}^{+0.01}$&$0.3_{-0.1}^{+0.1}$&$0.3_{-0.1}^{+0.1}$ & 1.05/266 \\\\
13&$15.3_{-1.8}^{+1.9}$&$0.8_{-0.15}^{+0.01}$&$22.1_{-1.3}^{+1.5}$&$11.9_{-0.4}^{+0.4}$&$0.43_{-0.06}^{+0.11}$&$25.3_{-1.7}^{+1.8}$&$5.4_{-1.1}^{+1.1}$&$9.4_{-0.7}^{+1.1}$&$1.1_{-0.05}^{+0.05}$&$0.07_{-0.01}^{+0.01}$&$0.4_{-0.1}^{+0.1}$&$0.3_{-0.1}^{+0.1}$ & 1.01/230 \\\\
14&$13.7_{-2.4}^{+2.4}$&$0.8_{-0.07}^{+0.05}$&$23.7_{-0.7}^{+0.7}$&$12.7_{-0.5}^{+0.6}$&$0.42_{-0.09}^{+0.08}$&$26.1_{-1.1}^{+0.7}$&$4.5_{-0.8}^{+0.8}$&$9.3_{-0.7}^{+0.9}$&$1.1_{-0.05}^{+0.05}$&$0.07_{-0.01}^{+0.01}$&$0.4_{-0.1}^{+0.1}$&$0.3_{-0.1}^{+0.1}$ & 1.02/281 \\\\
15&$15.9_{-1.2}^{+1.9}$&$0.8_{-0.05}^{+0.04}$&$19.1_{-1.8}^{+1.8}$&$12.3_{-1.5}^{+0.4}$&$0.33_{-0.14}^{+0.14}$&$25.6_{-3.7}^{+2.3}$&$4.4_{-0.5}^{+0.2}$&$8.9_{-0.6}^{+2.7}$&$0.9_{-0.05}^{+0.05}$&$0.05_{-0.01}^{+0.01}$&$0.4_{-0.1}^{+0.1}$&$0.3_{-0.1}^{+0.1}$ & 1.05/291 \\\\
16&$19.6_{-1.6}^{+1.7}$&$0.9_{-0.02}^{+0.02}$&$18.9_{-1.4}^{+1.4}$&$11.9_{-0.5}^{+0.5}$&$0.61_{-0.09}^{+0.07}$&$25.5_{-1.1}^{+1.5}$&$5.7_{-0.9}^{+0.4}$&$8.1_{-1.4}^{+2.4}$&$0.9_{-0.13}^{+0.09}$&$0.05_{-0.01}^{+0.01}$&$0.4_{-0.1}^{+0.1}$&$0.3_{-0.1}^{+0.1}$ & 1.09/154 \\\\
\hline
 \end{tabular}
\RaggedRight \\ \textbf{Notes: }${}^a$ Additional hydrogen column density in units $10^{22}$ \linebreak ${}^b$ 1 to 10 keV Flux in units of $10^{-10}$ ergs  cm$^{-2}$ s$^{-1}$ \linebreak${}^c$ In units of keV, Iron emission line equivalent width in units of $10^{-1}$
\end{table*}

\begin{table*}
\centering
\small
\setlength{\tabcolsep}{6.3pt} 
\renewcommand{\arraystretch}{0.85} 
\caption{The best fit parameters with NuSTAR data for Cen X-3. Model: tbabs*(pcfabs*gabs*newhcut*powerlaw + gaussian). Hydrogen column density for Interstellar absorption has been fixed at $0.64 \times 10^{22}$, width of secondary absorption fixed at 1.59 keV and  of fundamental at 5.01 keV, cutoff energy is fixed at 13.95 keV and smoothing width at 5 keV. Errors are within 90 \% confidence.}
\label{nu_table1}
\begin{tabular}{cccccccccccc} 
 \toprule
   \textbf{Sgmt} &\multicolumn{2}{c}{\textbf{pcfabs}}& \textbf{Unabs} & \multicolumn{2}{c}{\textbf{gabs {weak}}} & \multicolumn{2}{c}{\textbf{gabs {fund}}} & \textbf{newhcut} & \multicolumn{2}{c}{\textbf{powerlaw}}  
 &\textbf{$\chi^{2}_{red}$}\\
      \cmidrule(l){2-3}\cmidrule(l){5-6}\cmidrule(l){7-8}\cmidrule(l){10-11}\cmidrule(l){12-12}
       \textbf{No.} & $NH^a$ & $C_{fr}$ &\textbf{$Flux^b$} & $E0^c$ & $str^c$ & $E0^c$ & $str^c$ & $E_{fold}^{c}$  & $\Gamma$ & $norm^{d}$  & $d.o.f$\\ 
\bottomrule
&&&&&&&&&&&\\
\multicolumn{12}{l}{\textbf{First Binary Orbit: }} \\
&&&&&&&&&&&\\
1&$41.2_{-1.8}^{+1.1}$&$0.5_{-0.1}^{+0.1}$&$26.1_{-1.1}^{+1.2}$&$14.9_{-0.6}^{+0.6}$&$0.64_{-0.28}^{+0.34}$&$29.2_{-0.8}^{+1.3}$&$5.5_{-0.9}^{+0.8}$&$7.5_{-1.5}^{+0.4}$&$0.5_{-0.33}^{+0.35}$&$0.04_{-0.01}^{+0.01}$ & 1.05/1035 \\\\
2&$20.7_{-3.4}^{+3.7}$&$0.6_{-0.1}^{+0.1}$&$35.7_{-0.8}^{+0.9}$&$14.3_{-0.5}^{+0.5}$&$0.28_{-0.09}^{+0.11}$&$28.6_{-0.7}^{+1.1}$&$4.6_{-0.3}^{+0.5}$&$6.2_{-0.4}^{+0.6}$&$0.7_{-0.17}^{+0.21}$&$0.07_{-0.02}^{+0.02}$ & 1.01/1028 \\\\
3&$60.8_{-3.7}^{+2.3}$&$0.5_{-0.1}^{+0.1}$&$28.7_{-1.1}^{+1.1}$&$14.6_{-0.4}^{+0.4}$&$0.37_{-0.12}^{+0.13}$&$28.8_{-1.4}^{+1.3}$&$5.9_{-0.8}^{+0.8}$&$5.6_{-0.3}^{+0.4}$&$0.6_{-0.17}^{+0.19}$&$0.04_{-0.01}^{+0.01}$ & 1.01/1000 \\\\
4&$20.3_{-7.2}^{+6.7}$&$0.7_{-0.1}^{+0.1}$&$42.6_{-1.1}^{+1.3}$&$14.1_{-0.4}^{+0.4}$&$0.21_{-0.04}^{+0.04}$&$28.2_{-0.6}^{+0.9}$&$4.4_{-0.1}^{+0.2}$&$7.3_{-0.9}^{+1.5}$&$0.8_{-0.18}^{+0.22}$&$0.09_{-0.03}^{+0.03}$ & 1.01/1027 \\\\
5&$36.8_{-5.8}^{+3.8}$&$0.6_{-0.2}^{+0.1}$&$39.8_{-5.7}^{+2.1}$&$14.2_{-0.6}^{+1.1}$&$0.26_{-0.04}^{+0.04}$&$28.3_{-0.3}^{+0.4}$&$4.3_{-0.6}^{+0.2}$&$8.9_{-1.4}^{+1.4}$&$0.9_{-0.27}^{+0.17}$&$0.11_{-0.06}^{+0.07}$ & 1.02/1019 \\\\
6&$21.6_{-4.9}^{+3.8}$&$0.6_{-0.1}^{+0.1}$&$42.5_{-0.7}^{+0.7}$&$14.2_{-0.3}^{+0.3}$&$0.23_{-0.04}^{+0.04}$&$28.7_{-0.1}^{+0.5}$&$4.7_{-0.5}^{+0.2}$&$7.7_{-0.7}^{+0.8}$&$0.8_{-0.13}^{+0.11}$&$0.09_{-0.02}^{+0.03}$ & 1.02/1198 \\\\
7&$58.3_{-4.2}^{+4.4}$&$0.4_{-0.1}^{+0.2}$&$28.1_{-1.8}^{+3.5}$&$14.7_{-0.5}^{+0.4}$&$0.42_{-0.21}^{+0.22}$&$29.1_{-0.5}^{+1.2}$&$5.6_{-0.7}^{+0.4}$&$7.1_{-0.9}^{+1.3}$&$0.5_{-0.17}^{+0.23}$&$0.03_{-0.01}^{+0.03}$ & 1.04/1001 \\\\
8&$29.4_{-4.4}^{+3.6}$&$0.3_{-0.1}^{+0.1}$&$39.5_{-0.7}^{+0.8}$&$14.3_{-0.5}^{+0.6}$&$0.26_{-0.11}^{+0.12}$&$28.7_{-0.7}^{+1.1}$&$4.1_{-0.6}^{+0.9}$&$8.5_{-0.6}^{+0.7}$&$0.8_{-0.09}^{+0.09}$&$0.08_{-0.02}^{+0.02}$ & 1.01/1216 \\\\
9&$72.9_{-5.6}^{+4.7}$&$0.4_{-0.1}^{+0.1}$&$27.9_{-0.9}^{+1.2}$&$14.7_{-0.8}^{+0.7}$&$0.51_{-0.11}^{+0.11}$&$28.9_{-0.6}^{+0.7}$&$4.8_{-0.3}^{+0.4}$&$7.4_{-0.7}^{+0.9}$&$0.7_{-0.09}^{+0.12}$&$0.04_{-0.01}^{+0.02}$ & 1.02/1043 \\\\
10&$56.8_{-3.1}^{+4.3}$&$0.3_{-0.1}^{+0.1}$&$29.6_{-1.1}^{+1.1}$&$14.5_{-0.7}^{+0.9}$&$0.35_{-0.11}^{+0.11}$&$28.8_{-0.6}^{+0.9}$&$5.2_{-0.4}^{+0.1}$&$5.7_{-0.4}^{+0.7}$&$0.5_{-0.02}^{+0.01}$&$0.03_{-0.01}^{+0.02}$ & 1.04/951 \\\\
11&$59.9_{-4.3}^{+3.9}$&$0.6_{-0.1}^{+0.1}$&$30.3_{-0.9}^{+2.8}$&$14.5_{-0.5}^{+0.6}$&$0.31_{-0.11}^{+0.11}$&$28.8_{-0.1}^{+0.8}$&$5.5_{-0.9}^{+0.3}$&$9.3_{-1.9}^{+1.9}$&$0.7_{-0.08}^{+0.03}$&$0.09_{-0.03}^{+0.02}$ & 1.02/898 \\\\
12&$17.9_{-3.8}^{+4.9}$&$0.6_{-0.1}^{+0.1}$&$39.9_{-1.1}^{+1.1}$&$14.2_{-0.6}^{+1.4}$&$0.24_{-0.06}^{+0.08}$&$28.6_{-0.4}^{+0.6}$&$4.3_{-0.1}^{+0.3}$&$7.1_{-0.9}^{+1.1}$&$0.8_{-0.18}^{+0.19}$&$0.07_{-0.03}^{+0.01}$ & 1.01/1031 \\\\
13&$15.1_{-4.6}^{+4.9}$&$0.4_{-0.1}^{+0.1}$&$70.1_{-1.3}^{+1.4}$&$13.8_{-0.7}^{+0.8}$&$0.11_{-0.03}^{+0.12}$&$27.5_{-0.1}^{+0.9}$&$3.9_{-0.8}^{+0.1}$&$8.9_{-0.8}^{+1.1}$&$1.3_{-0.09}^{+0.08}$&$0.04_{-0.01}^{+0.01}$ & 1.04/1059 \\\\
14&$26.6_{-4.6}^{+2.4}$&$0.7_{-0.1}^{+0.1}$&$50.7_{-1.1}^{+1.2}$&$13.9_{-0.2}^{+0.2}$&$0.21_{-0.04}^{+0.04}$&$28.1_{-0.9}^{+1.6}$&$4.3_{-0.3}^{+0.4}$&$8.4_{-0.9}^{+0.9}$&$1.1_{-0.15}^{+0.06}$&$0.02_{-0.01}^{+0.01}$ & 1.02/1032 \\\\
15&$28.6_{-3.9}^{+3.3}$&$0.7_{-0.2}^{+0.1}$&$30.4_{-0.9}^{+1.1}$&$14.4_{-0.5}^{+0.6}$&$0.31_{-0.15}^{+0.15}$&$28.7_{-0.7}^{+1.1}$&$4.8_{-0.3}^{+0.2}$&$6.8_{-0.6}^{+1.4}$&$0.7_{-0.29}^{+0.25}$&$0.05_{-0.02}^{+0.01}$ & 1.02/1012 \\\\
&&&&&&&&&&&\\
\multicolumn{12}{l}{\textbf{Second Binary Orbit: }} \\
&&&&&&&&&&&\\
16&$21.3_{-4.6}^{+4.5}$&$0.8_{-0.1}^{+0.1}$&$48.1_{-1.1}^{+1.2}$&$13.9_{-0.3}^{+0.3}$&$0.21_{-0.04}^{+0.04}$&$28.1_{-0.5}^{+0.6}$&$4.1_{-0.6}^{+0.7}$&$6.9_{-0.4}^{+1.2}$&$1.1_{-0.17}^{+0.15}$&$0.17_{-0.06}^{+0.05}$ & 1.05/1126 \\\\
17&$7.2_{-1.4}^{+1.3}$&$0.4_{-0.2}^{+0.1}$&$78.4_{-0.8}^{+0.9}$&$13.7_{-0.5}^{+0.5}$&$0.11_{-0.02}^{+0.01}$&$27.2_{-0.2}^{+1.2}$&$3.8_{-0.4}^{+0.3}$&$7.9_{-0.4}^{+0.5}$&$1.2_{-0.05}^{+0.06}$&$0.41_{-0.05}^{+0.07}$ & 1.07/1249 \\\\
18&$3.2_{-1.3}^{+1.7}$&$0.3_{-0.1}^{+0.1}$&$93.7_{-0.5}^{+0.6}$&$13.5_{-0.4}^{+0.4}$&$0.05_{-0.03}^{+0.06}$&$27.1_{-1.1}^{+1.3}$&$3.3_{-0.7}^{+0.4}$&$6.1_{-0.2}^{+0.2}$&$1.3_{-0.22}^{+0.03}$&$0.42_{-0.01}^{+0.01}$ & 1.07/1249 \\\\
19&$4.1_{-1.3}^{+1.8}$&$0.4_{-0.1}^{+0.1}$&$88.9_{-0.2}^{+1.1}$&$13.6_{-0.8}^{+0.6}$&$0.11_{-0.04}^{+0.06}$&$27.2_{-1.1}^{+0.9}$&$3.1_{-0.3}^{+0.6}$&$9.2_{-0.5}^{+0.6}$&$1.2_{-0.01}^{+0.01}$&$0.41_{-0.01}^{+0.01}$ & 1.02/1178 \\\\
20&$6.2_{-1.4}^{+1.7}$&$0.4_{-0.1}^{+0.2}$&$85.4_{-0.6}^{+0.7}$&$13.7_{-0.3}^{+0.3}$&$0.16_{-0.13}^{+0.15}$&$27.5_{-1.3}^{+0.9}$&$3.6_{-0.5}^{+0.4}$&$8.1_{-0.4}^{+0.5}$&$1.2_{-0.01}^{+0.01}$&$0.47_{-0.01}^{+0.01}$ & 1.02/1180 \\\\
21&$7.6_{-1.8}^{+1.3}$&$0.6_{-0.2}^{+0.1}$&$85.1_{-0.5}^{+0.5}$&$13.8_{-0.4}^{+0.4}$&$0.05_{-0.03}^{+0.06}$&$27.5_{-0.1}^{+0.9}$&$3.5_{-0.6}^{+0.6}$&$8.7_{-0.5}^{+0.7}$&$1.2_{-0.01}^{+0.05}$&$0.52_{-0.01}^{+0.01}$ & 1.04/1137 \\\\
22&$17.9_{-1.5}^{+1.8}$&$0.6_{-0.1}^{+0.2}$&$58.5_{-0.7}^{+0.7}$&$14.1_{-0.4}^{+0.4}$&$0.21_{-0.06}^{+0.06}$&$27.8_{-0.2}^{+0.2}$&$4.1_{-0.1}^{+0.2}$&$8.1_{-0.5}^{+0.6}$&$1.2_{-0.07}^{+0.07}$&$0.26_{-0.05}^{+0.06}$ & 1.06/1195 \\\\
23&$57.3_{-3.1}^{+2.8}$&$0.6_{-0.1}^{+0.2}$&$30.6_{-0.8}^{+0.7}$&$14.4_{-0.2}^{+0.3}$&$0.27_{-0.21}^{+0.11}$&$29.1_{-0.5}^{+0.4}$&$5.3_{-0.7}^{+0.6}$&$5.4_{-0.6}^{+0.7}$&$0.3_{-0.23}^{+0.2}$&$0.02_{-0.01}^{+0.01}$ & 1.07/1004 \\\\
&&&&&&&&&&&\\
\hline\\
\multicolumn{12}{r}{\textbf{ Continued to Table ~\ref{tab:nu_line}}} \\
 \end{tabular}
\RaggedRight \linebreak \textbf{Notes: }${}^a$ Additional Hydrogen Column density in units $10^{22}$ \linebreak ${}^b$ 3 to 10 keV Flux in units of $10^{-10}$ ergs  cm$^{-2}$ s$^{-1}$ \linebreak${}^c$ In units of keV \linebreak${}^d$ Powerlaw normalisation and errors in units of $10^{-1}$
\label{tab:nu_fit}
\end{table*}

\begin{table*}
\centering
\small
\setlength{\tabcolsep}{16pt} 
\renewcommand{\arraystretch}{0.9} 
\caption{The best fit parameters with NuSTAR data for Cen X-3 Model: tbabs*(pcfabs*gabs*newhcut*powerlaw + gaussian). Iron emission line energies fixed at 5.7 keV, 6.4 keV and 6.9 keV. Errors are within 90 \% confidence.}
\label{nu_table2}
\begin{tabular}{cccccccc} 
\multicolumn{8}{r}{\textbf{Continued from Table ~\ref{tab:nu_fit}}} \\
&&&&&&&\\
 \toprule
    \textbf{Sgmt} &
      \multicolumn{2}{c}{\textbf{5.7 keV line}} & \multicolumn{2}{c}{\textbf{6.4 keV line}} &  \multicolumn{2}{c}{\textbf{6.9 keV line}}  &\multicolumn{1}{c}{\textbf{$\chi^{2}_{red}$}}\\
      \cmidrule(l){2-3}\cmidrule(l){4-5}\cmidrule(l){6-7}\cmidrule(l){8-8}
       \textbf{No.} & $Flux^a$ & $Eqw^b$ &  $Flux^a$ & $Eqw^b$ & $Flux^a$ & $Eqw^b$ & $d.o.f$\\ 
\bottomrule
&&&&&&&\\
&&&&&&&\\
\multicolumn{8}{l}{\textbf{First Binary Orbit: }} \\
&&&&&&&\\
1	& $0.11_{-0.01}^{+0.02}$ & $0.19_{-0.13}^{+0.28}$ & $0.22_{-0.02}^{+0.02}$ & $0.3_{-0.26}^{+0.34}$ & $0.09_{-0.01}^{+0.01}$ & $0.05_{-0.04}^{+0.06}$	  & $ 1.05/1035 $	\\\\
2	& $0.09_{-0.04}^{+0.04}$ & $0.19_{-0.12}^{+0.28}$ & $0.29_{-0.04}^{+0.05}$ & $0.2_{-0.16}^{+0.23}$ & $0.08_{-0.03}^{+0.03}$ & $0.06_{-0.03}^{+0.08}$	 & $ 1.01/1028 $	\\\\
3	& $0.19_{-0.09}^{+0.08}$ & $0.09_{-0.06}^{+0.14}$ & $0.26_{-0.05}^{+0.04}$ & $0.3_{-0.25}^{+0.35}$ & $0.09_{-0.04}^{+0.03}$ & $0.07_{-0.06}^{+0.08}$	 & $ 1.01/1000 $	\\\\
4	& $0.13_{-0.11}^{+0.09}$ & $0.09_{-0.02}^{+0.17}$ & $0.37_{-0.06}^{+0.05}$ & $0.2_{-0.17}^{+0.23}$ & $0.06_{-0.04}^{+0.04}$ & $0.04_{-0.01}^{+0.06}$	 & $ 1.01/1027 $	\\\\
5	& $0.12_{-0.09}^{+0.09}$ & $0.09_{-0.02}^{+0.19}$ & $0.39_{-0.07}^{+0.07}$ & $0.3_{-0.25}^{+0.36}$ & $0.08_{-0.05}^{+0.04}$ & $0.05_{-0.02}^{+0.09}$	 & $ 1.02/1019 $	\\\\
6	& $0.21_{-0.06}^{+0.06}$ & $0.09_{-0.07}^{+0.13}$ & $0.38_{-0.03}^{+0.03}$ & $0.2_{-0.18}^{+0.22}$ & $0.09_{-0.03}^{+0.03}$ & $0.05_{-0.04}^{+0.06}$	 & $ 1.02/1198 $	\\\\
7	& $0.15_{-0.06}^{+0.04}$ & $0.19_{-0.12}^{+0.26}$ & $0.39_{-0.04}^{+0.04}$ & $0.4_{-0.36}^{+0.44}$ & $0.08_{-0.03}^{+0.02}$ & $0.08_{-0.05}^{+0.09}$	 & $ 1.04/1001 $	\\\\
8	& $0.26_{-0.05}^{+0.05}$ & $0.19_{-0.16}^{+0.24}$ & $0.38_{-0.03}^{+0.03}$ & $0.2_{-0.19}^{+0.21}$ & $0.09_{-0.02}^{+0.02}$ & $0.05_{-0.04}^{+0.06}$	 & $ 1.01/1216 $	\\\\
9	& $0.08_{-0.03}^{+0.02}$ & $0.09_{-0.07}^{+0.13}$ & $0.38_{-0.03}^{+0.03}$ & $0.4_{-0.37}^{+0.43}$ & $0.06_{-0.02}^{+0.02}$ & $0.06_{-0.04}^{+0.08}$	 & $ 1.02/1043 $	\\\\
10	& $0.19_{-0.07}^{+0.05}$ & $0.19_{-0.17}^{+0.23}$ & $0.39_{-0.06}^{+0.06}$ & $0.3_{-0.28}^{+0.32}$ & $0.09_{-0.04}^{+0.03}$ & $0.07_{-0.06}^{+0.08}$	 & $ 1.04/951 $	\\\\
11	& $0.02_{-0.01}^{+0.05}$ & $0.09_{-0.01}^{+0.06}$ & $0.32_{-0.05}^{+0.05}$ & $0.3_{-0.26}^{+0.34}$ & $0.04_{-0.02}^{+0.03}$ & $0.03_{-0.02}^{+0.05}$	 & $ 1.02/898 $	\\\\
12	& $0.24_{-0.09}^{+0.09}$ & $0.19_{-0.12}^{+0.28}$ & $0.31_{-0.06}^{+0.06}$ & $0.2_{-0.17}^{+0.24}$ & $0.09_{-0.04}^{+0.04}$ & $0.05_{-0.04}^{+0.07}$	 & $ 1.01/1031 $	\\\\
13	& $0.16_{-0.09}^{+0.12}$ & $0.05_{-0.03}^{+0.09}$ & $0.54_{-0.06}^{+0.08}$ & $0.1_{-0.07}^{+0.13}$ & $0.06_{-0.01}^{+0.05}$ & $0.05_{-0.01}^{+0.02}$	 & $ 1.04/1059 $	\\\\
14	& $0.09_{-0.06}^{+0.07}$ & $0.04_{-0.02}^{+0.09}$ & $0.23_{-0.05}^{+0.06}$ & $0.1_{-0.08}^{+0.13}$ & $0.01_{-0.01}^{+0.02}$ & $0.05_{-0.02}^{+0.79}$	 & $ 1.02/1032 $	\\\\
15	& $0.14_{-0.08}^{+0.08}$ & $0.04_{-0.02}^{+0.06}$ & $0.18_{-0.01}^{+0.05}$ & $0.2_{-0.15}^{+0.26}$ & $0.06_{-0.04}^{+0.04}$ & $0.05_{-0.02}^{+0.08}$	 & $ 1.02/1012 $	\\\\
&&&&&&&\\
\multicolumn{8}{l}{\textbf{Second Binary Orbit: }} \\
&&&&&&&\\
16	& $0.11_{-0.09}^{+0.09}$ & $0.06_{-0.01}^{+0.12}$ & $0.16_{-0.06}^{+0.06}$ & $0.1_{-0.06}^{+0.13}$ & $0.03_{-0.01}^{+0.04}$ & $0.01_{-0.01}^{+0.03}$	 & $ 1.05/1126 $	\\\\
17	& $0.16_{-0.12}^{+0.12}$ & $0.07_{-0.05}^{+0.09}$ & $0.32_{-0.04}^{+0.04}$ & $0.1_{-0.09}^{+0.11}$ & $0.02_{-0.01}^{+0.01}$ & $0.03_{-0.02}^{+0.04}$	 & $ 1.07/1249 $	\\\\
18	& $0.35_{-0.04}^{+0.04}$  & $0.05_{-0.04}^{+0.07}$ & $0.56_{-0.06}^{+0.06}$ & $0.1_{-0.09}^{+0.11}$ & $0.09_{-0.02}^{+0.02}$ & $0.02_{-0.02}^{+0.02}$	 & $ 1.07/1249 $	\\\\
19	& $0.26_{-0.09}^{+0.07}$ & $0.08_{-0.07}^{+0.09}$ & $0.46_{-0.03}^{+0.04}$ & $0.1_{-0.09}^{+0.11}$ & $0.05_{-0.03}^{+0.03}$ & $0.02_{-0.01}^{+0.02}$	 & $ 1.02/1178 $	\\\\
20	& $0.24_{-0.06}^{+0.05}$ & $0.09_{-0.08}^{+0.09}$ & $0.39_{-0.04}^{+0.05}$ & $0.1_{-0.08}^{+0.11}$ & $0.03_{-0.02}^{+0.01}$ & $0.02_{-0.01}^{+0.02}$	 & $ 1.02/1180 $	\\\\
21	& $0.13_{-0.06}^{+0.06}$ & $0.04_{-0.03}^{+0.06}$ & $0.44_{-0.05}^{+0.05}$ & $0.1_{-0.08}^{+0.11}$ & $0.01_{-0.01}^{+0.04}$ & $0.03_{-0.01}^{+0.07}$	 & $ 1.04/1137 $	\\\\
22	& $0.16_{-0.07}^{+0.07}$ & $0.07_{-0.04}^{+0.09}$ & $0.31_{-0.04}^{+0.04}$ & $0.1_{-0.09}^{+0.11}$ & $0.03_{-0.03}^{+0.03}$ & $0.02_{-0.01}^{+0.02}$	 & $ 1.06/1195 $	\\\\
23	& $0.12_{-0.09}^{+0.09}$ & $0.13_{-0.02}^{+0.24}$ & $0.24_{-0.08}^{+0.08}$ & $0.2_{-0.16}^{+0.25}$ & $0.04_{-0.02}^{+0.05}$ & $0.03_{-0.01}^{+0.07}$	 & $ 1.07/1004 $	\\\\
&&&&&&&\\
\bottomrule\\
\end{tabular}

\RaggedRight \textbf{Notes: }${}^a$ 3 to 10 keV Flux in units of $10^{-10}$ ergs  cm$^{-2}$ s$^{-1}$ \linebreak${}^b$ In units of keV
\label{tab:nu_line}
\end{table*}

 \section{Discussions}
 \label{sec:discuss}
\vspace{5pt}
\begin{figure}
    \centering
    \begin{subfigure}[b]{0.36\textwidth}
        \centering
            \includegraphics[width=\textwidth]{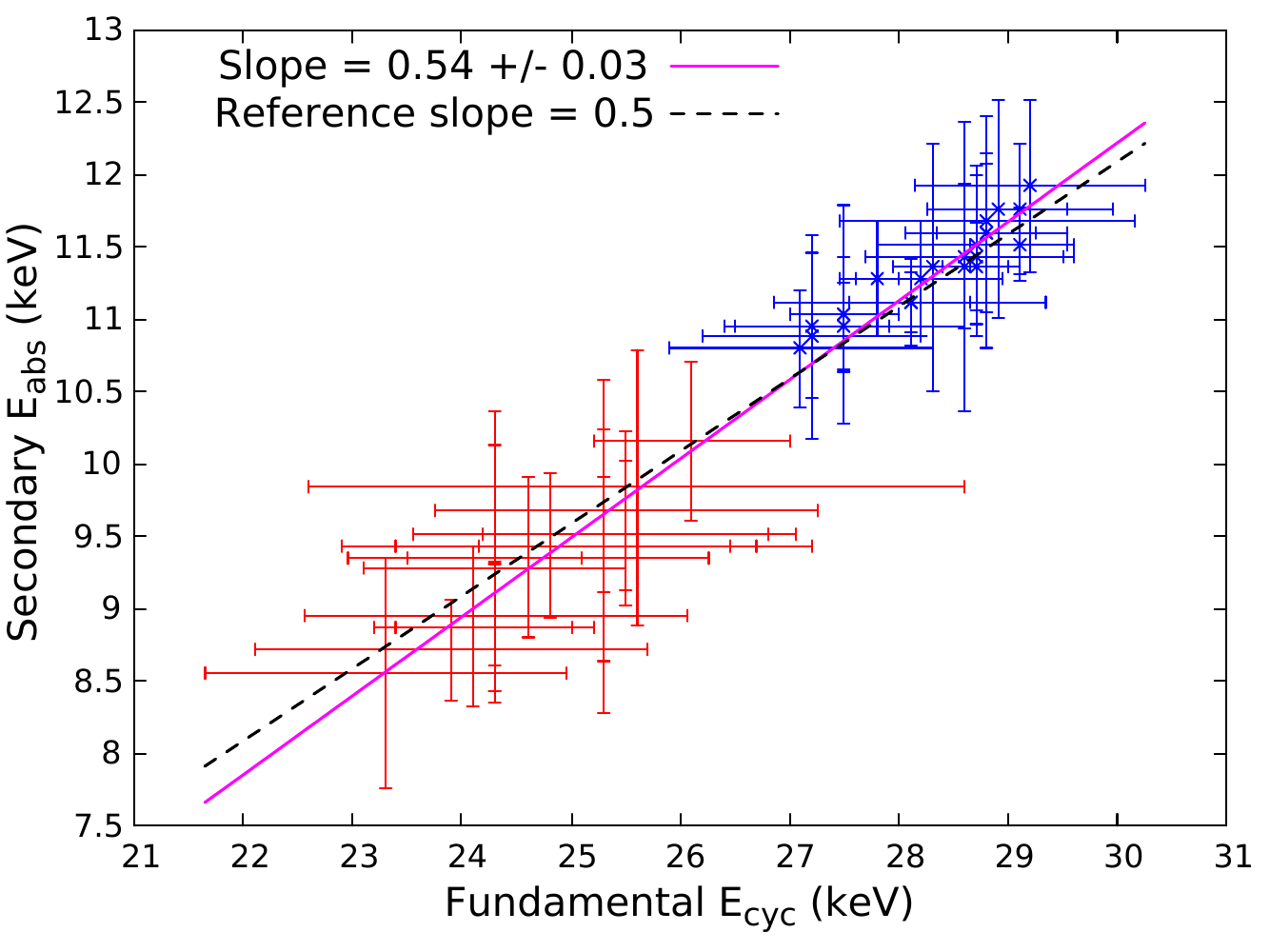}
            \caption{}
            \label{fig:linecorr1}
    \end{subfigure}
    \begin{subfigure}[b]{0.36\textwidth}
        \centering
        \includegraphics[width=\textwidth]{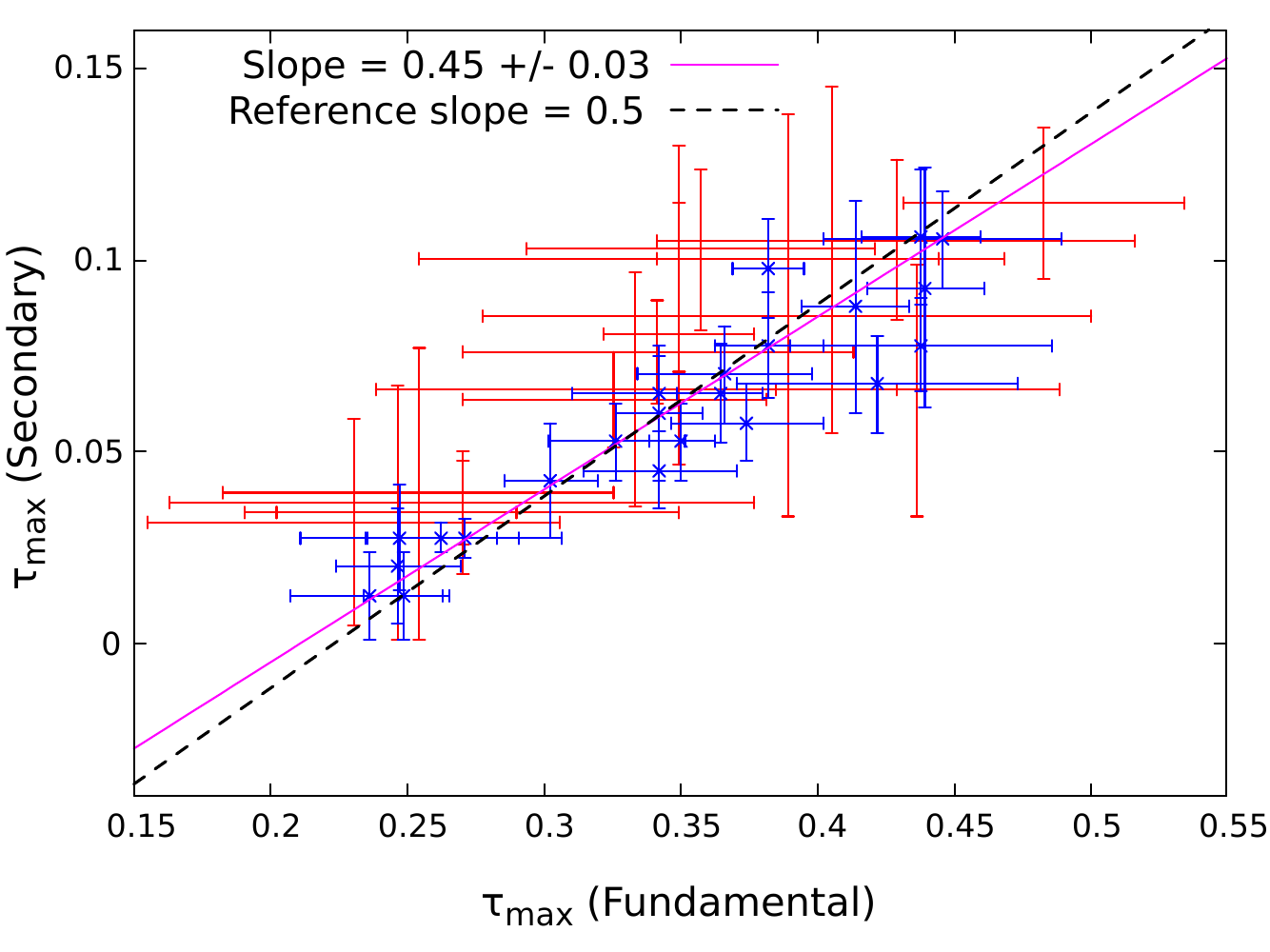}
        \caption{}
        \label{fig:linecorr2}
    \end{subfigure}
    \caption{\fontsize{7}{20} \selectfont { Joint Correlation of secondary absorption feature with fundamental CRSF feature using ASTROSAT and NuSTAR data. \textbf{a}: Secondary absorption energy vs fundamental CRSF absorption energy. The blue points in the upper group represent data from NuSTAR, while the red points in the lower group represent data from ASTROSAT. Common line with a slope 0.54 $\pm$ 0.03 fits both the observations. The slope gives the ratio of weak energy to fundamental line energy. The dashed line is the reference line for slope 0.5 for weak energy to be exactly half of fundamental energy. The slope of correlations falls within 2$\sigma$ deviation from the reference slope. \textbf{b}: Secondary absorption optical depth showing a positive correlation to fundamental CRSF optical depth. Points marked in red are from ASTROSAT data and in blue are from NuSTAR data. Slope 0.45  $\pm$  0.03 suggests secondary absorption depth to be $\sim$ 45\% of fundamental.}}
    \label{fig:linecorr}
\end{figure}

In the ASTROSAT observation of March 8, 2018 (covering almost a full orbit) and NuSTAR observation of January 12, 2022 (covering two intensity state in two orbital phases) light curve showed high variability in count rate over the orbital phase with dips and bumps in non eclipse regions. We have adopted time resolved spectral analysis of Cen X-3. We divided the light curve into various time segments and carried individual spectral analysis. Initially we implemented the phenomenological model to 1 to 30 keV spectra of ASTROSAT which provided a reasonable fit to the X-ray continuum. However, a weak absorption feature $\sim$ 11.56 keV was seen. We modified the model and adopted two gaussian absorption components for weak and the fundamental cyclotron absorption which provided a fit to all the segments with $\chi^{2}_{red} \sim$ 1. Similar feature was observed with NuSTAR at energy $\sim$ 14.48 keV. We extended this model to 3 to 30 keV spectra of NuSTAR with an addition of two more gaussian emission components. 6.4 keV line emission, a broad line emission $\sim$ 5.7 keV and a narrow line emission $\sim$ 6.87 keV was seen in spectra with NuSTAR. 

Time resolved segments in Cen X-3 showed significant variation in luminosity in both observations. ASTROSAT observation showed variation between 1.2 and 5.3 in units of $10^{37}$ ergs s$^{-1}$ and that in NuSTAR observation between 1.6 and 5.8 in units of $10^{37}$ ergs s$^{-1}$. Distance to source is taken to be 7.2 kpc \citep{vall22}. Cen X-3 luminosities in all time segments in both the observations are well above critical luminosity ($\sim 10^{37}$ ergs s$^{-1}$). We studied correlation of cyclotron line energy and optical depth for both fundamental and secondary absorption and found a non-linear negative correlation trend represented by Equation \ref{eq:eq1} with correlation index $\beta \sim$  0.08.

There has been more than 35 X-ray pulsars that has CRSF in their X-ray spectrum and many has shown either correlation or anti-correlation of CRSF centroid energy with luminosity. Sources like V0332+53 \citep{Tsygankov2010} and SMC X-2 \citep{jaisawal2023} has shown anti-correlation. \citet{basko1975} proposed the presence of a radiative shock in the accretion column near the neutron star surface for very high luminous sources when the accretion becomes super Eddington above critical luminosity of $\sim$ $10^{37}$ ergs s$^{-1}$. \citet{Poutanen_2013} has pointed out that in a dipole field, the magnetic field strength change with the radiative shock height is much larger. Dependence of CRSF line energy with varying accretion rate may be due to change of local geometry or structures in magnetic field. In-falling plasma is confined by the magnetic field at the polar cap of neutron star and accretion mound is formed. This produces an effect of distortion of local magnetic field from dipolar magnetic field \citep{nishimura2005}. Cyclotron resonance in distorted magnetic field may lead to lowering of CRSF energy as well as multiple absorption features with anharmonicity \citep{nishimura2011}. There has been some sources reported (V0332+53, 4U 1538-52, A 0535+26 etc) to have multiple absorption features with anharmonicity which is understood to be due to the above distortion of local field.

\citet{dipanjan2012} obtained mound structure by solving the Grad-Shafranov equation and simulated the cyclotron spectra originating from the mound surface. The resultant spectra was generated by integrating the emission from multiple parts of the mound and performed a pulse phase resolved analysis. Assuming mound height of 60 metres, their simulation showed the presence of redshifted cyclotron feature as a small dip $\sim$ 8 keV along with the fundamental CRSF $\sim$ 18 keV due to distorted field. It was inferred by authors that the CRSF line energy may not always represent the dipole magnetic field. 

In an attempt to explore the relation of this weak absorption to fundamental CRSF, we studied the correlation of optical depth and energy between two absorptions and found a linear positive correlations. We jointly fitted the correlation plot between secondary (Y-axis) and fundamental (X-axis) absorption energies from all segments from both the missions with a straight line (Figure \ref{fig:linecorr1}), the slope of which corresponds to the energy ratio and made a comparison of the best fit slope to that of a reference sub-harmonic line having slope 0.5 which would correspond to case where the secondary absorption energy is exactly at half of the fundamental CRSF energy. We followed similar procedure for depth (Figure \ref{fig:linecorr2}).
Weak absorption energy to fundamental absorption energy ratio is found to be 0.54 $\pm$ 0.03 and their corresponding depth ratio to be 0.45 $\pm$ 0.03. The deviation of best fit line from sub-harmonic line is within 2$\sigma$ confidence for both depth and energy.

The negative correlation between the cyclotron line optical depth and flux (Bottom panel of Figure \ref{fig:astrofund} and Figure \ref{fig:nucorr2}) may be a consequence of the changing conditions in the accretion flow at different accretion rates. Higher accretion rates lead to a decrease in cyclotron line optical depth due to increased plasma effects and possible screening of the magnetic field leads to a shift in the cyclotron line energy to lower values, resulting in a stronger X-ray flux. Anti-correlation of CRSF energy and depth to flux and simultaneously detecting a secondary absorption in our work comes in support of the resonance taking place with the distorted magnetic field due to high accretion rate. In addition to fundamental CRSF, a weak absorption $\sim$ 11.6 keV (ASTROSAT) and $\sim$ 14.5 keV (NuSTAR) reported in this work can be speculated to be the redshifted dipolar cyclotron resonance feature exhibiting sub-harmonic behaviour.

\section{Conclusions}
\label{sec:conclusion}
\vspace{5pt}
We carried a time resolved spectroscopy of Cen X-3 with ASTROSAT observation of 2018, March 8 to 10 and NuSTAR observations of 2022, January 12 to 16. Results revealed a weak secondary absorption $\sim$ 1.6 keV wide with centroid energy at almost half the value of the fundamental CRSF centroid energy. Depth of secondary absorption was $\sim$ 45 \% of fundamental absorption depth. Fundamental cyclotron line energy and flux showed anti-correlation for the first time related by a normalised powerlaw trend with negative index of 0.08 $\pm$ 0.01. The secondary absorption energy showed similar trend to flux. Depth of both fundamental and secondary absorption also showed similar trend but with higher negative correlation index $\sim$ 0.4 with ASTROSAT and $\sim$ 0.5 with NuSTAR data. We suggest this feature to have same source of origin as that of fundamental CRSF and possibly could be the redshifted dipolar cyclotron resonance feature exhibiting sub-harmonic behaviour.

\section*{Acknowledgements}
\vspace{5pt}
We are thankful to the anonymous referee for providing valuable suggestions. P. Dangal and NKC extend their thanks to the Director and Dean of IUCAA, Pune, for the local hospitality and facilities provided.
We have used the publicly available data from the ASTROSAT mission of the Indian Space Research Organization (ISRO), archived at the Indian Space Science Data Centre (ISSDC) and NASA's High Energy Astrophysics Science Archive Research Center (HEASARC). This work has made use of the data processing and analysis packages of ASTROSAT mission and HEASOFT version 6.29. We have made use of the NASA Astrophysics Data System (NASA ADS). All plots, other than spectra, are made in \texttt{gnuplot} \citep{gnuplot}.\\
\vspace{-20pt}
\section*{Data Availability}
\vspace{5pt}
The data used in this paper are publicly available at the ASTROSAT data archive (\url{https://astrobrowse.issdc.gov.in/astro_archive/archive/Home.jsp}) and NASA's HEASARC (\url{https://heasarc.gsfc.nasa.gov/cgi-bin/W3Browse/w3browse.pl}.

\vspace{-15pt}

\bibliographystyle{mnras}
\bibliography{ref}

\begin{thebibliography}{}
\makeatletter
\relax
\def\mn@urlcharsother{\let\do\@makeother \do\$\do\&\do\#\do\^\do\_\do\%\do\~}
\def\mn@doi{\begingroup\mn@urlcharsother \@ifnextchar [ {\mn@doi@}
  {\mn@doi@[]}}
\def\mn@doi@[#1]#2{\def\@tempa{#1}\ifx\@tempa\@empty \href
  {http://dx.doi.org/#2} {doi:#2}\else \href {http://dx.doi.org/#2} {#1}\fi
  \endgroup}
\def\mn@eprint#1#2{\mn@eprint@#1:#2::\@nil}
\def\mn@eprint@arXiv#1{\href {http://arxiv.org/abs/#1} {{\tt arXiv:#1}}}
\def\mn@eprint@dblp#1{\href {http://dblp.uni-trier.de/rec/bibtex/#1.xml}
  {dblp:#1}}
\def\mn@eprint@#1:#2:#3:#4\@nil{\def\@tempa {#1}\def\@tempb {#2}\def\@tempc
  {#3}\ifx \@tempc \@empty \let \@tempc \@tempb \let \@tempb \@tempa \fi \ifx
  \@tempb \@empty \def\@tempb {arXiv}\fi \@ifundefined
  {mn@eprint@\@tempb}{\@tempb:\@tempc}{\expandafter \expandafter \csname
  mn@eprint@\@tempb\endcsname \expandafter{\@tempc}}}

\bibitem[\protect\citeauthoryear{{Ash}, {Reynolds}, {Roche}, {Norton}, {Still}
  \& {Morales-Rueda}}{{Ash} et~al.}{1999}]{asht99}
{Ash} T.~D.~C.,  {Reynolds} A.~P.,  {Roche} P.,  {Norton} A.~J.,  {Still}
  M.~D.,   {Morales-Rueda} L.,  1999, \mn@doi [\mnras]
  {10.1046/j.1365-8711.1999.02605.x}, \href
  {https://ui.adsabs.harvard.edu/abs/1999MNRAS.307..357A} {307, 357}

\bibitem[\protect\citeauthoryear{{Audley}}{{Audley}}{1998}]{audley98}
{Audley} M.~D.,  1998, {A Broad-band Spectral and Timing Study of the X-Ray
  Binary System Centaurus X-3}, NASA/CR-1998-206896; NAS 1.26:206896, PhD
  thesis

\bibitem[\protect\citeauthoryear{{Bachhar}, {Raman}, {Bhalerao}  \&
  {Bhattacharya}}{{Bachhar} et~al.}{2022}]{bachhar2022}
{Bachhar} R.,  {Raman} G.,  {Bhalerao} V.,   {Bhattacharya} D.,  2022, \mn@doi
  [\mnras] {10.1093/mnras/stac2901}, \href
  {https://ui.adsabs.harvard.edu/abs/2022MNRAS.517.4138B} {517, 4138}

\bibitem[\protect\citeauthoryear{{Basko} \& {Sunyaev}}{{Basko} \&
  {Sunyaev}}{1975}]{basko1975}
{Basko} M.~M.,  {Sunyaev} R.~A.,  1975, \aap, \href
  {https://ui.adsabs.harvard.edu/abs/1975A&A....42..311B} {42, 311}

\bibitem[\protect\citeauthoryear{{Becker} \& {Wolff}}{{Becker} \&
  {Wolff}}{2005}]{beck05}
{Becker} P.~A.,  {Wolff} M.~T.,  2005, \mn@doi [\apj] {10.1086/431720}, \href
  {https://ui.adsabs.harvard.edu/abs/2005ApJ...630..465B} {630, 465}

\bibitem[\protect\citeauthoryear{{Burderi}, {Di Salvo}, {Robba}, {La Barbera}
  \& {Guainazzi}}{{Burderi} et~al.}{2000}]{burderi2000}
{Burderi} L.,  {Di Salvo} T.,  {Robba} N.~R.,  {La Barbera} A.,   {Guainazzi}
  M.,  2000, \mn@doi [\apj] {10.1086/308336}, \href
  {https://ui.adsabs.harvard.edu/abs/2000ApJ...530..429B} {530, 429}

\bibitem[\protect\citeauthoryear{{Ebisawa}, {Day}, {Kallman}, {Nagase},
  {Kotani}, {Kawashima}, {Kitamoto}  \& {Woo}}{{Ebisawa} et~al.}{1996}]{ebis96}
{Ebisawa} K.,  {Day} C. S.~R.,  {Kallman} T.~R.,  {Nagase} F.,  {Kotani} T.,
  {Kawashima} K.,  {Kitamoto} S.,   {Woo} J.~W.,  1996, \mn@doi [\pasj]
  {10.1093/pasj/48.3.425}, \href
  {https://ui.adsabs.harvard.edu/abs/1996PASJ...48..425E} {48, 425}

\bibitem[\protect\citeauthoryear{{Gaia Collaboration} et~al.,}{{Gaia
  Collaboration} et~al.}{2023}]{vall22}
{Gaia Collaboration} et~al., 2023, \mn@doi [\aap]
  {10.1051/0004-6361/202243940}, \href
  {https://ui.adsabs.harvard.edu/abs/2023A&A...674A...1G} {674, A1}

\bibitem[\protect\citeauthoryear{{Giacconi}, {Gursky}, {Kellogg}, {Schreier}
  \& {Tananbaum}}{{Giacconi} et~al.}{1971}]{giac71}
{Giacconi} R.,  {Gursky} H.,  {Kellogg} E.,  {Schreier} E.,   {Tananbaum} H.,
  1971, \mn@doi [\apjl] {10.1086/180762}, \href
  {https://ui.adsabs.harvard.edu/abs/1971ApJ...167L..67G} {167, L67}

\bibitem[\protect\citeauthoryear{{Harrison} et~al.,}{{Harrison}
  et~al.}{2013}]{harrison2013}
{Harrison} F.~A.,  et~al., 2013, \mn@doi [\apj] {10.1088/0004-637X/770/2/103},
  \href {https://ui.adsabs.harvard.edu/abs/2013ApJ...770..103H} {770, 103}

\bibitem[\protect\citeauthoryear{{Hutchings}, {Cowley}, {Crampton}, {van
  Paradijs}  \& {White}}{{Hutchings} et~al.}{1979}]{Hutchings1979}
{Hutchings} J.~B.,  {Cowley} A.~P.,  {Crampton} D.,  {van Paradijs} J.,
  {White} N.~E.,  1979, \mn@doi [\apj] {10.1086/157042}, \href
  {https://ui.adsabs.harvard.edu/abs/1979ApJ...229.1079H} {229, 1079}

\bibitem[\protect\citeauthoryear{{Jaisawal} et~al.,}{{Jaisawal}
  et~al.}{2023}]{jaisawal2023}
{Jaisawal} G.~K.,  et~al., 2023, \mn@doi [\mnras] {10.1093/mnras/stad781},
  \href {https://ui.adsabs.harvard.edu/abs/2023MNRAS.521.3951J} {521, 3951}

\bibitem[\protect\citeauthoryear{{Kallman} \& {White}}{{Kallman} \&
  {White}}{1989}]{killman1989}
{Kallman} T.,  {White} N.~E.,  1989, \mn@doi [\apj] {10.1086/167554}, \href
  {https://ui.adsabs.harvard.edu/abs/1989ApJ...341..955K} {341, 955}

\bibitem[\protect\citeauthoryear{{Krzeminski}}{{Krzeminski}}{1974}]{krzemenski1974}
{Krzeminski} W.,  1974, \mn@doi [\apjl] {10.1086/181609}, \href
  {https://ui.adsabs.harvard.edu/abs/1974ApJ...192L.135K} {192, L135}

\bibitem[\protect\citeauthoryear{{Mukherjee} \& {Bhattacharya}}{{Mukherjee} \&
  {Bhattacharya}}{2012}]{dipanjan2012}
{Mukherjee} D.,  {Bhattacharya} D.,  2012, \mn@doi [\mnras]
  {10.1111/j.1365-2966.2011.20085.x}, \href
  {https://ui.adsabs.harvard.edu/abs/2012MNRAS.420..720M} {420, 720}

\bibitem[\protect\citeauthoryear{{Nagase}}{{Nagase}}{1989}]{naga89}
{Nagase} F.,  1989, \pasj, \href
  {https://ui.adsabs.harvard.edu/abs/1989PASJ...41....1N} {41, 1}

\bibitem[\protect\citeauthoryear{{Naik} \& {Paul}}{{Naik} \&
  {Paul}}{2012}]{Naik}
{Naik} S.,  {Paul} B.,  2012, Bulletin of the Astronomical Society of India,
  \href {https://ui.adsabs.harvard.edu/abs/2012BASI...40..503N} {40, 503}

\bibitem[\protect\citeauthoryear{{Nishimura}}{{Nishimura}}{2011}]{nishimura2011}
{Nishimura} O.,  2011, \mn@doi [\apj] {10.1088/0004-637X/730/2/106}, \href
  {https://ui.adsabs.harvard.edu/abs/2011ApJ...730..106N} {730, 106}

\bibitem[\protect\citeauthoryear{{Nishimura}}{{Nishimura}}{2005}]{nishimura2005}
{Nishimura} O.,  2011,2005, \mn@doi [\pasj] {10.1093/pasj/57.5.769}, \href
  {https://ui.adsabs.harvard.edu/abs/2005PASJ...57..769N} {57, 769}

\bibitem[\protect\citeauthoryear{{Poutanen}, {Mushtukov}, {Suleimanov},
  {Tsygankov}, {Nagirner}, {Doroshenko}  \& {Lutovinov}}{{Poutanen}
  et~al.}{2013}]{Poutanen_2013}
{Poutanen} J.,  {Mushtukov} A.~A.,  {Suleimanov} V.~F.,  {Tsygankov} S.~S.,
  {Nagirner} D.~I.,  {Doroshenko} V.,   {Lutovinov} A.~A.,  2013, \mn@doi
  [\apj] {10.1088/0004-637X/777/2/115}, \href
  {https://ui.adsabs.harvard.edu/abs/2013ApJ...777..115P} {777, 115}

\bibitem[\protect\citeauthoryear{{Raichur} \& {Paul}}{{Raichur} \&
  {Paul}}{2008}]{raichur2008}
{Raichur} H.,  {Paul} B.,  2008, \mn@doi [\apj] {10.1086/591037}, \href
  {https://ui.adsabs.harvard.edu/abs/2008ApJ...685.1109R} {685, 1109}

\bibitem[\protect\citeauthoryear{{Santangelo}, {del Sordo}, {Segreto}, {dal
  Fiume}, {Orlandini}  \& {Piraino}}{{Santangelo} et~al.}{1998}]{santa98}
{Santangelo} A.,  {del Sordo} S.,  {Segreto} A.,  {dal Fiume} D.,  {Orlandini}
  M.,   {Piraino} S.,  1998, \aap, \href
  {https://ui.adsabs.harvard.edu/abs/1998A&A...340L..55S} {340, L55}

\bibitem[\protect\citeauthoryear{{Schreier}, {Levinson}, {Gursky}, {Kellogg},
  {Tananbaum}  \& {Giacconi}}{{Schreier} et~al.}{1972}]{schreier1972evidence}
{Schreier} E.,  {Levinson} R.,  {Gursky} H.,  {Kellogg} E.,  {Tananbaum} H.,
  {Giacconi} R.,  1972, \mn@doi [\apjl] {10.1086/180896}, \href
  {https://ui.adsabs.harvard.edu/abs/1972ApJ...172L..79S} {172, L79}

\bibitem[\protect\citeauthoryear{{Singh} et~al.,}{{Singh}
  et~al.}{2014}]{kpsingh2014}
{Singh} K.~P.,  et~al., 2014, in {Takahashi} T.,  {den Herder} J.-W.~A.,
  {Bautz} M.,  eds,  SPIE Conference Series Vol. 9144, Space Telescopes and
  Instrumentation 2014: Ultraviolet to Gamma Ray. p. 91441S,
  \mn@doi{10.1117/12.2062667}

\bibitem[\protect\citeauthoryear{{Singh} et~al.,}{{Singh}
  et~al.}{2017}]{singh2017soft}
{Singh} K.~P.,  et~al., 2017, \mn@doi [Journal of Astrophysics and Astronomy]
  {10.1007/s12036-017-9448-7}, \href
  {https://ui.adsabs.harvard.edu/abs/2017JApA...38...29S} {38, 29}

\bibitem[\protect\citeauthoryear{{Staubert} et~al.,}{{Staubert}
  et~al.}{2019}]{staubert2019}
{Staubert} R.,  et~al., 2019, \mn@doi [\aap] {10.1051/0004-6361/201834479},
  \href {https://ui.adsabs.harvard.edu/abs/2019A&A...622A..61S} {622, A61}

\bibitem[\protect\citeauthoryear{{Suchy} et~al.,}{{Suchy}
  et~al.}{2008}]{suchy2008}
{Suchy} S.,  et~al., 2008, \mn@doi [\apj] {10.1086/527042}, \href
  {https://ui.adsabs.harvard.edu/abs/2008ApJ...675.1487S} {675, 1487}

\bibitem[\protect\citeauthoryear{{Takeshima}, {Dotani}, {Mitsuda}  \&
  {Nagase}}{{Takeshima} et~al.}{1991}]{take91}
{Takeshima} T.,  {Dotani} T.,  {Mitsuda} K.,   {Nagase} F.,  1991, \pasj, \href
  {https://ui.adsabs.harvard.edu/abs/1991PASJ...43L..43T} {43, L43}

\bibitem[\protect\citeauthoryear{{Tananbaum}, {Gursky}, {Kellogg}, {Levinson},
  {Schreier}  \& {Giacconi}}{{Tananbaum} et~al.}{1972}]{tana72}
{Tananbaum} H.,  {Gursky} H.,  {Kellogg} E.~M.,  {Levinson} R.,  {Schreier} E.,
    {Giacconi} R.,  1972, \mn@doi [\apjl] {10.1086/180968}, \href
  {https://ui.adsabs.harvard.edu/abs/1972ApJ...174L.143T} {174, L143}

\bibitem[\protect\citeauthoryear{{Thompson} \& {Rothschild}}{{Thompson} \&
  {Rothschild}}{2009}]{thom09}
{Thompson} T. W.~J.,  {Rothschild} R.~E.,  2009, \mn@doi [\apj]
  {10.1088/0004-637X/691/2/1744}, \href
  {https://ui.adsabs.harvard.edu/abs/2009ApJ...691.1744T} {691, 1744}

\bibitem[\protect\citeauthoryear{{Tomar}, {Pradhan}  \& {Paul}}{{Tomar}
  et~al.}{2021}]{tomar2021}
{Tomar} G.,  {Pradhan} P.,   {Paul} B.,  2021, \mn@doi [\mnras]
  {10.1093/mnras/staa3477}, \href
  {https://ui.adsabs.harvard.edu/abs/2021MNRAS.500.3454T} {500, 3454}

\bibitem[\protect\citeauthoryear{{Tsygankov}, {Lutovinov}  \&
  {Serber}}{{Tsygankov} et~al.}{2010}]{Tsygankov2010}
{Tsygankov} S.~S.,  {Lutovinov} A.~A.,   {Serber} A.~V.,  2010, \mn@doi
  [\mnras] {10.1111/j.1365-2966.2009.15791.x}, \href
  {https://ui.adsabs.harvard.edu/abs/2010MNRAS.401.1628T} {401, 1628}

\bibitem[\protect\citeauthoryear{Williams, Kelley  \& {many others}}{Williams
  et~al.}{2010}]{gnuplot}
Williams T.,  Kelley C.,   {many others} 2010, Gnuplot 4.4: an interactive
  plotting program, \url{http://gnuplot.sourceforge.net/}

\bibitem[\protect\citeauthoryear{{Yadav} et~al.,}{{Yadav}
  et~al.}{2016}]{yadav2016}
{Yadav} J.~S.,  et~al., 2016, in {den Herder} J.-W.~A.,  {Takahashi} T.,
  {Bautz} M.,  eds,  SPIE Conference Series Vol. 9905, Space Telescopes and
  Instrumentation 2016: Ultraviolet to Gamma Ray. p. 99051D,
  \mn@doi{10.1117/12.2231857}

\bibitem[\protect\citeauthoryear{{Yang}, {Wang}, {Liu}, {Chen}, {Wu}, {Tian}
  \& {Chen}}{{Yang} et~al.}{2023}]{wang2023}
{Yang} W.,  {Wang} W.,  {Liu} Q.,  {Chen} X.,  {Wu} H.~J.,  {Tian} P.~F.,
  {Chen} J.~S.,  2023, \mn@doi [\mnras] {10.1093/mnras/stad048}, \href
  {https://ui.adsabs.harvard.edu/abs/2023MNRAS.519.5402Y} {519, 5402}

\makeatother
\end{thebibliography}

\bsp	
\label{lastpage}
\end{document}